\documentclass[aps,amsfonts,nofootinbib]{revtex4-1}
\usepackage{caption}
\usepackage{subcaption}
\usepackage{epsfig}
\usepackage{graphicx}
\usepackage{slashed}
\usepackage{amsmath}
\usepackage{amsbsy}
\usepackage{bm}
\usepackage{color}
\usepackage{epsfig}
\newcommand{\ee}{\end{equation}}
\newcommand{\bb}{\begin{equation}}
\newcommand{\eqb}{\begin{eqnarray}}
\newcommand{\eqf}{\end{eqnarray}}

\newcommand{\1}{{\'{\i}}}

\def\1{\'{\i}}

\def\1{\'{\i}}

\begin{document}

\title{Cosmological Kinetic Mixing }

\author{Ashok Das }
%\email{@gmail.com}
\affiliation{$^{a}$ Department of Physics and Astronomy, University of Rochester, Rochester, NY 14627-0171, USA}
\affiliation{$^{b}$ Saha Institute of Nuclear Physics, 1/AF Bidhannagar, Calcutta 700064, India}

\author{Jorge  Gamboa}
%\email{@gmail.com}
\affiliation{Departamento de F\1sica, Universidad de Santiago de Chile, Casilla 307, Santiago, Chile}

\author{Miguel  Pino}
%\email{}
\affiliation{Departamento de F\1sica, Universidad de Santiago de Chile, Casilla 307, Santiago, Chile}

\begin{abstract}
In this paper we generalize the kinetic mixing idea to time reparametrization invariant theories, namely, relativistic point particles and cosmology in order to obtain new insights for dark matter and energy. In the first example, two relativistic particles interact through an appropriately chosen coupling term. It is shown that the system can be diagonalized by means of a non-local field redefinition, and, as a result of this procedure, the mass of one the particles gets rescaled. In the second case, inspired by the previous example, two cosmological models (each with its own scale factor) are made to interact in a similar fashion. The equations of motion are solved numerically in different scenarios (dust, radiation or a cosmological constant coupled to each sector of the system). When a cosmological constant term is present, kinetic mixing rescales it to a lower value which may be more amenable to observations.

\end{abstract}

\pacs{PACS numbers: 98.80.Cq,98.80.-k,98.80.Jk}
\date{\today}
\maketitle

\section{Introduction}\label{intro}

The understanding of the dark matter/energy problem is one of the great challenges of modern particle physics and cosmology, and it has been studied from different perspectives in the last years \cite{review1}. 

An interesting advance in this direction --at least in the context of dark matter--  is based on the interaction between standard (visible) photons and  ``dark'' or ``hidden'' photons which couple to visible matter only through  kinetic mixing \cite{holdom}. Relics of this mixing can manifest through explicit relations between the coupling constants and other observables (such as Sommerfeld enhancement) \cite{papers,nos}. 

Although the experimental search for hidden sectors has not given positive results so far \cite{dobrich}, it is still a fertile arena for proposing physics beyond the standard model and kinetic mixing is an appealing mechanism to introduce new physics. 

A relevant and important question to answer in this context is that of a generalization of kinetic mixing to the case of gravity. Namely, whether we can use kinetic mixing as a guide line for constructing modified gravity theories with some interesting phenomenological consequences such as, for example, a natural explanation for dark energy. Although we do not have an answer to this question yet, the objective of this work is to go forward in this direction, trying to understand how this mechanism works and if this modification of gravity can be useful in understanding the dark energy problem from a different perspective. 

To the best of our knowledge, presently there are no kinetic mixing mechanisms for gravity, although the closest idea is bigravity which arises from a different point of view (for example, by including interactions between metrics as extensions of massive gravity  \cite{damour}). Since gravity is a non-linear and covariant theory, to build such a  mechanism is not easy. However, we take a modest step by considering a relativistic massless particle as the most simple example of a generally covariant system. We build the covariant analogue of kinetic mixing in this case, where the gauge group is spanned by worldline time reparametrization. Then, we move forward to a cosmological model (a minisuperspace of General Relativity (GR) with FRW metric), which presents several similarities with the relativistic particle. As shown below in this paper a kinetic mixing mechanism can be build in this case leading to some interesting phenomenological implications.

The paper is organized as follows. In the next section we briefly consider the example of photons to explain kinetic mixing. In section \ref{km_particles}, we generalize this idea to relativistic particles and study some of its implications. In section \ref{km_cosm}, the results are extended to cosmology with some of its implications, while section \ref{conclusions} contains a  discussion of our results. 

 \section{Motivating with an example}\label{motivation}
         
 In order to explain the main idea of kinetic mixing, let us consider the Lagrangian density 
 \eqb  \label{lag}
 {\cal L} &=& {\bar \psi} \left( i {\partial \hspace{-.5em}  \slash  \hspace{.10em}}- {eA \hspace{-.5em}  \slash  \hspace{.10em}} \right) \psi - 
\frac{1}{4}F_{\mu\nu} (A) F^{\mu\nu} (A) - 
 \frac{1}{4 }F_{\mu\nu} (B) F^{\mu\nu} (B) + \frac{\gamma}{2} F_{\mu\nu} (A) F^{\mu\nu} (B) , 
 \eqf
 where $\psi$ is a fermion field coupled to the standard photon field $A_\mu$ with coupling constant $e$. The field $B_\mu$ is a hidden $U(1)$ gauge field coupled only to $A_\mu$ through the kinetic mixing term 
 $$ \frac{\gamma}{2} F_{\mu\nu}(A) F^{\mu\nu}(B), $$
 where $\gamma$ is a dimensionless coupling constant. Here $F_{\mu\nu}(A), F_{\mu\nu}(B)$ are the field strength tensors for the photon field and the hidden gauge field respectively. This coupling was introduced in the eighties by Holdom \cite{holdom} as a way to propose physics beyond the standard model. 
 
 A simple diagonalization of the system in \eqref{lag} can be carried out as follows. Let us redefine  
 \begin{equation}
   B_\mu = \tilde B_\mu+ \gamma A_\mu ,
 \end{equation}
which allows us to write 
 \eqb 
 \mathcal{L} = {\bar \psi} \left( i {\partial \hspace{-.5em}  \slash  \hspace{.10em}}- {eA \hspace{-.5em}  \slash  \hspace{.10em}} \right) \psi -  \frac{1}{4}(1-\gamma^2)  F_{\mu\nu} (A) F^{\mu\nu} (A)- \frac{1}{4}F_{\mu\nu} (\tilde B) F^{\mu\nu} (\tilde B) \label{dia}.
 \eqf
The extra multiplicative factor in the $F^2(A)$ term can be absorbed into a redefinition of the electric charge $e$ by rescaling the field $A_\mu$ as 
\begin{equation}
  A_\mu \rightarrow \frac{1}{\sqrt{1-\gamma^2}}A_\mu,
\end{equation}
giving
 \eqb 
 \mathcal{L} = {\bar \psi} \left( i {\partial \hspace{-.5em}  \slash  \hspace{.10em}}- {\tilde e A \hspace{-.5em}  \slash  \hspace{.10em}} \right) \psi -  \frac{1}{4} F_{\mu\nu} (A) F^{\mu\nu} (A) - \frac{1}{4}F_{\mu\nu} (\tilde B) F^{\mu\nu} (\tilde B), \label{dia}
 \eqf
 where the modified charge $\tilde e$ is defined to be 
 \bb 
 \tilde e = \frac{e}{\sqrt{1 -\gamma^2}}. \label{relation}
 \ee
 Therefore, $A_\mu$ and $\tilde B_\mu$ are decoupled and the only relic of kinetic mixing is  in the rescaling of the electric charge \eqref{relation}.

\section{Kinetic Mixing for Relativistic Particles}\label{km_particles} 

In this section we generalize the mechanism of kinetic mixing to relativistic particles. First, let us briefly review the Lagrangian description for the free relativistic particles . As is well known, the standard Lagrangian for the massive relativistic point  particle is given by 
\bb\label{rel}
L[x] =  m \sqrt{{\dot x}^\mu {\dot x}_\mu},
\ee 
where $x^\mu$ is the position four vector of the particle, $m$ its mass and we use the Bjorken-Drell metric. The dot stands for a derivative with respect to the proper time parameter $\tau$.

An equivalent formulation, more suitable for our purposes, is obtained by introducing an auxiliary field $N$, 
\bb \label{relN}
L[x,N]=  \frac{1}{2 N} {\dot x}^\mu {\dot x}_\mu + \frac{m^2}{2}N.
\ee 
Both Lagrangian  are equivalent in the sense that they give rise to the same equations of motion. Furthermore, the Lagrangian \eqref{rel} can be obtained from \eqref{relN} by eliminating the auxiliary field $N$ using its equation of motion. Formulation \eqref{relN} has the advantage of having a well defined massless limit. 
  
The equation of motion for $x_\mu$ is 
\bb 
 \frac{d}{d\tau}\left( \frac{{\dot x}^\mu}{N}\right) =0, 
\ee 
implying that $\frac{{\dot x}^\mu}{N}$ is a constant. On the other hand, the equation of motion of $N$ gives the constraint 
\bb 
-\frac{{\dot x}^\mu {\dot x}_\mu}{N^2} + m^2 =0,
\ee
which describes the relativistic dispersion relation.  
 
Note that the Lagrangian \eqref{relN} possesses a gauge invariance, namely,  time reparametrization invariance. The field transformations are given by
\begin{equation}
    \delta x^\mu= \epsilon \dot{x}^\mu\;\;\;\;\;\;\delta N = \frac{d}{d\tau}(\epsilon N),
\end{equation}
where $\epsilon$ is an infinitesimal  function of $\tau$.
\\

Since we are interested in extending the kinetic mixing mechanism to generally covariant systems, we will start considering the most simple example, namely, two massless relativistic particles as discussed in the last section, described by the action 
\bb 
S = \int d \tau \left( \frac{1}{2N} {\dot x}_\mu {\dot x}^\mu + \frac{1}{2M} {\dot y}_\mu {\dot y}^\mu + \frac{\gamma}{\sqrt{NM}} {\dot x}_\mu {\dot y}^\mu \right),  \label{part1}
\ee 
where $x^\mu$ and $y^\mu$ are the position four vectors of the two particles, $N$ and $M$ are auxiliary fields and $\gamma$ is a coupling constant. The choice of the interaction is dictated by invariance under time reparametrizations\footnote{The interaction of two relativistic particles has been studied previously in a different context in \cite{gomis}}.

Just as in the electromagnetic case \eqref{lag}, the action \eqref{part1} can be diagonalized by performing a field redefinition. However, in this case, this redefinition turns out to be non-local. In order to see this, we rewrite the action as 
\bb 
S = \int d\tau\left(\frac{(1-\gamma^2)}{2N} {\dot x}^2  + \frac{1}{2M}\left(  {\dot y}_\mu+\gamma \sqrt{\frac{M}{N}}\dot{x}_\mu \right)^2\right),
\ee 
where $x^2\equiv x_\mu x^\mu$. By performing the non-local redefinition
\begin{equation}\label{redef}
  \dot z_\mu\equiv \dot y_\mu +\gamma\sqrt{ \frac{M}{N}} \dot x_\mu,
\end{equation}
the action can be written as
\begin{equation}\label{diagonal1}
  S = \int  d\tau \left(\frac{(1-\gamma^2)}{2N} {\dot x}^2  + \frac{1}{2M} {\dot z}^2\right),
\end{equation}
which describes two decoupled  relativistic particles. Note that $\gamma$ must be chosen different from $1$,  otherwise, action \eqref{part1} describes only one relativistic particle.    

Two comments are in order. First, note that we could have included a mass term for the $x$-particle
\begin{equation}
  S = \int d\tau\left(\frac{(1-\gamma^2)}{2N} {\dot x}^2 +\frac{m^2}{2}N + \frac{1}{2M} {\dot z}^2\right). 
\end{equation}
The extra factor $(1-\gamma^2)$ can be absorbed binto a rescaling of the mass $m$. To see this, we redefine $N$ as
\begin{equation}
  N\rightarrow (1-\gamma^2) N,
\end{equation}
obtaining
\begin{equation}
  S = \int d\tau\left(\frac{1}{2N} {\dot x}^2 +\frac{\tilde m^2}{2}N + \frac{1}{2M} {\dot z}^2\right),
\end{equation}
where $\tilde m$ is defined as
\begin{equation}
  \tilde m = m \sqrt{1-\gamma^2}.
\end{equation}
This indicates that, as with the charge $e$ in the electromagnetic case, the effect of the kinetic mixing of two relativistic particles is a rescaling of the mass of one of the particles. (Note that we can only add a mass term for one of the particles. A mass term for both particles would spoil the diagonalization procedure.) 

A second interesting comment is that the diagonal form of the action \eqref{diagonal1} makes explicit the gauge symmetries of the system, which corresponds to two independent time reparametrizations. Indeed, the infinitesimal transformations are
\begin{align}
    \delta x^\mu= \epsilon_1 \dot{x}^\mu,\;\;\;\;\;\;&\delta N = \frac{d}{d\tau}(\epsilon_1 N),\notag\\
  \delta z^\mu= \epsilon_2 \dot{z}^\mu,\;\;\;\;\;\;&\delta M = \frac{d}{d\tau}(\epsilon_2 M),
\end{align}
or, in terms of the original fields we have 
\begin{align}\label{symmetries}
  \delta x^\mu&= \epsilon_1 \dot{x}^\mu,\notag\\
  \delta \dot y_\mu &=  \frac{d}{d\tau}(\epsilon_2 \dot y_\mu)+\frac{\gamma}{2}\sqrt{\frac{M}{N^3}}\dot N (\epsilon_1-\epsilon_2)\dot x_\mu -\gamma\sqrt{\frac{M}{N}}(\epsilon_1-\epsilon_2)\ddot x_\mu -\frac{\gamma}{2}\sqrt{\frac{M}{N}}(\dot \epsilon_1-\dot \epsilon_2)\dot x_\mu,\notag\\
  \delta N &= \frac{d}{d\tau}(\epsilon_1 N),\notag\\
  \delta M &= \frac{d}{d\tau}(\epsilon_2 M).
\end{align}
Note that, since the field redefinition \eqref{redef} is non-local, the infinitesimal transformation for $\delta \dot y_\mu$ is rather nontrivial. These two symmetries are useful when studying the equations of motion since they allow for the gauge choice $N=M=1$.

\section{Cosmological Implications}\label{km_cosm}

In this section we will analyze the implications of our results when applied to cosmology. If we assume the spatial curvature of the universe equal to zero ($k=0$), the  Friedmann equations read
\eqb  
&&\left( \frac{{\dot a} }{a} \right)^2 =   \frac{8\pi G}{3}\rho, \label{fr1}
\\
&& 2\frac{{\ddot a}}{a} +\left(\frac{\dot{a} }{a}\right)^2+8\pi G p = 0, \label{fr2}
\eqf
where $a(t)$ is the cosmological scale factor, while $\rho(t)$ and $p(t)$ are the energy density and pressure of matter. $G$ stands for Newton's constant. 

In order to study kinetic mixing in this case, it is useful to obtain Friedmann equations from an action principle. Of course, such an action exists. It is given by the Einstein-Hilbert Lagrangian
\begin{equation}
  I=\int\left( -\frac{1}{16 \pi G} \sqrt{-g}R + \sqrt{-g}L_M\right),
\end{equation}
where $g_{\mu\nu}$ is the space-time metric and $R$ the Ricci scalar. $L_M$ is a matter Lagrangian that leads to $\rho$ and $p$ in the Friedmann equations. Varying with respect to the metric, we obtain Einstein's equations
\begin{equation}\label{einstein}
  R_{\mu\nu}-\frac{1}{2}R g_{\mu\nu}=8\pi G \left(2 \frac{\delta L_M}{\delta g^{\mu\nu}}-g_{\mu\nu}L_M \right),
\end{equation}
where we can identify the right hand side with the energy-momentum tensor 
\begin{equation}
T_{\mu\nu}\equiv 2 \frac{\delta L_M}{\delta g^{\mu\nu}}-g_{\mu\nu}L_M.
\end{equation}

To obtain Friedmann's equations, we must use the ansatz for the metric  
\begin{equation}\label{ansatz}
  ds^2= N^2 dt^2-a^2 (dx^2+dy^2+dz^2),
\end{equation}
where $N$ and $a$ are functions only of time $t$, while the energy-momentum tensor is chosen to be that of a {\bf comoving } perfect fluid
\begin{equation}\label{perfect}
  T_{\mu\nu}=(p+\rho)U_\mu U_\nu -p g_{\mu\nu}.
\end{equation}
The vector $U_\mu$ stands for the fluid 4-velocity. Since the fluid is comoving, the spatial components of $U^{\mu}$ vanish. Besides, it fulfills the condition $U_\mu U^{\mu}=1$. These observations completely fix $U_\mu$. Plugging the ansatz \eqref{ansatz} and \eqref{perfect} into Einstein's equations \eqref{einstein}, we recover Friedmann equations \eqref{fr1} and \eqref{fr2}. 

Since we are interested in a simple and tractable case, without the complications associated with a fully covariant theory, we build a minisuperspace out of the Einstein-Hilbert action using the ansatz \eqref{ansatz} for the metric. Plugging the ansatz into the gravitational part of the action, we obtain the Lagrangian  
\bb 
L[a,N] = -3\frac{a\, {\dot a}^2}{N}, \label{lan}
\ee
whose equations of motion coincide with the Friedmann equations without sources. From now on, we set $8\pi G=1$. The curvature conventions are chosen such that $V_{\mu;\nu\rho}-V_{\mu;\rho\nu}=V_\alpha R^{\alpha}_{\;\;\mu\nu\rho}$. 

It is straightforward to check that the Lagrangian \eqref{lan} has a gauge invariance given by 
\begin{equation}\label{rep}
  \delta a = \epsilon \dot{a}, \quad\quad \delta N = \frac{d}{dt}(\epsilon N),
\end{equation}
which corresponds to time reparametrization. It is worth noting the similarity of this with the relativistic particle discussed earlier. If we include a cosmological constant, the Lagrangian has the form
\begin{equation}
L[a,N] = -3\frac{a\, {\dot a}^2}{N}-\Lambda a^3 N.
\end{equation} 
Redefining the field variables as $a=-A^{1/3}$ and $NA\to  N$, the Lagrangian can be written as 
\begin{equation}
L[A,N] = \frac{2}{3}\left(\frac{{\dot A}^2}{2 N}+\frac{3\Lambda}{2} N \right),
\end{equation}
which is precisely the Lagrangian for the one-dimensional massive  particle \eqref{relN}, where the cosmological constant plays the role of mass $m^2=3\Lambda$. Since the effect of kinetic mixing in the relativistic particle case corresponds to rescaling of the mass, we expect that in the case of the cosmological model, the consequence of this mechanism is to rescale the cosmological constant.

To include matter in the system \eqref{lan}, we need to supplement the action with the $\sqrt{-g}L_M$ term. The matter Lagrangian in this case defines a constrained system (because of the condition $U^\mu U_\mu=1$) and needs to be handled appropriately. Without going into technical details, it is enough for our purposes to note that, when varying the metric, the matter Lagrangian contributes with  
\begin{equation}\label{matter}
  \delta (\sqrt{-g} L_M)=3Na^2p\,\delta a-a^3\rho \delta N.
\end{equation}
The first Friedmann equation \eqref{fr1} ($G_0^0$ part of the Einstein's equations) comes from  the variation of the action with respect to $N$, while the second Friedmann equation ($G_i^i$)  comes from the Euler-Lagrange equation for $a$, combined with the earlier equation. Due to the gauge invariance \eqref{rep}, we can choose the condition $N=1$ in the equations.

The second Friedmann equation \eqref{fr2} can be cast in a more compact form. Taking the time derivative of  \eqref{fr1} and substituting the value of $\ddot{a}$ from equation into \eqref{fr2}, we obtain the equation
\begin{equation}\label{cons}
  \frac{d}{dt}(\rho a^3) =-p\frac{d}{dt}( a^3),
\end{equation}
which resembles the first law of thermodynamics.
 
Summarizing, the first Friedmann equation \eqref{fr1}, equation \eqref{cons} along with an equation of state $f(p,\rho)=0$, can be solved simultaneously to give the evolution of the scale factor $a$, the density $\rho$ and the pressure $p$ with time, fully determining the system.

\subsection{Including $a(t)$ and $b(t)$ Fields}

Having these facts in mind, the cosmological Lagrangian for two fields $a(t)$ and $b(t)$ with a kinetic mixing is 
\bb 
L[a,b,N,M] = -3\frac{a \,{\dot a}^2}{N} - 3\frac{b \,{\dot b}^2}{M} + 6\gamma\frac{\dot a \dot b}{\sqrt{N M}},  \label{min1}
\ee
where the kinetic mixing have been constructed in analogy with the two-relativistic particles discussed above.  

Note that the Lagrangian \eqref{min1} can also be diagonalized by means of a non-local redefinition of fields. By completing the square
\begin{equation}\label{action11}
L[a,b,N,M]=-3\frac{a \,{\dot a}^2}{N}\left(1-\frac{\gamma^2}{ab} \right) - 3\frac{b (\dot b -\gamma\sqrt{\frac{M}{N}}\frac{\dot a}{b})^2}{M},
\end{equation}
we can make the field redefinitions
\begin{equation}\label{redef11}
  \frac{N}{1-\frac{\gamma^2}{ab}} \to N, \quad\quad \dot b -\gamma\sqrt{\frac{M}{N}}\frac{\dot a}{b} \to \dot c, \quad\quad   \frac{M}{b} \to \frac{M}{c},
\end{equation}
leading to  
\begin{equation}
 L= -3\frac{a \,{\dot a}^2}{N} - 3\frac{c \,{\dot c}^2}{ M}
\end{equation}
which represent two non-interacting cosmological models. Accordingly, the Lagrangian \eqref{min1} has two time reparametrization symmetries, one for each independent sector. These transformations in terms of the fields $a$, $b$, $N$ and $M$ are non-local due to the redefinitions in \eqref{redef11}. See the appendix for a discussion on this point. However, it is important to remember that these two symmetries allow us to impose gauge fixing conditions on $N$ and $M$. 

We can calculate the equations of motion, including matter as in \eqref{matter}. The equation of motion for  $N$ and $M$ yield to 
\begin{align}
  \frac{\dot a^2}{a^2}=&\frac{\rho}{3} + \gamma \frac{\dot a \dot b}{a^3},\label{fri_km1}\\
 \frac{\dot b^2}{b^2}=&\frac{\tilde \rho}{3} + \gamma \frac{\dot a \dot b}{b^3}\label{fri_km2},
\end{align}
while the equations associated to $a$ and $b$ are
\begin{align}
  \dot a^2+2a\ddot a+a^2p-2\gamma\ddot b&=0,\\
  \dot b^2+2b\ddot b+b^2\tilde p-2\gamma\ddot a&=0.\
\end{align}
We are choosing $N=M=1$ from now on. The notation used is $\rho$ and $p$ for the density and pressure of the matter coupled to the $a$-sector, while $\tilde\rho$ and $\tilde p$ are the corresponding magnitudes in the $b$-sector. 

Using the same trick shown above, we can transform the last two equations into two first law equations. This gives
\begin{align}
  \frac{d}{dt} (\rho a^3 + 3\gamma\dot a \dot b )&= -(p -2\gamma\frac{\ddot b}{a^2})\frac{d}{dt}(a^3),\label{cons1}\\
   \frac{d}{dt} (\tilde \rho b^3 + 3\gamma\dot a \dot b)&=-(\tilde p-2\gamma\frac{\ddot a}{b^2})\frac{d}{dt}(b^3)\label{cons2}\,.
\end{align}
Note that the kinetic mixing can be interpreted as a time dependent contribution to the energy density and pressure on both sectors.

The set of Friedmann equations \eqref{fri_km1} and \eqref{fri_km2}, the equations \eqref{cons1} and \eqref{cons2}, along with two equations of state $f(p,\rho)=0$ and $\tilde f(\tilde p,\tilde \rho)=0$, completely determine the state of the system. Nonetheless, even for the simplest cases of matter (radiation or pressureless matter), an analytical solution is very hard to obtain. However, as shown in the next section, we can use numerical integration of the equations to study the behavior of the fields.

\subsection{Numerical Integration}

In this section, we choose some appropriated equations of state, $f(p,\rho)=0$ and $\tilde f(\tilde p,\tilde \rho)=0$, and solve equations \eqref{fri_km1}, \eqref{fri_km2}, \eqref{cons1} and \eqref{cons2} numerically. For simplicity, the equations of state will describe sources such as pressureless matter, radiation or a cosmological constant. It will be shown below that the most interesting case is when considering a cosmological constant coupled to one of the sectors.

The numerical integration of the equations is done with the arbitrary initial conditions $$a(0)=b(0)=1,\quad\quad \rho(0)=\tilde \rho(0)=1.$$ The overall behavior of the evolution of the system is not  affected by choosing different boundary conditions. Note that initial conditions for the first derivatives of the fields are not needed, since they can be expressed algebraically in terms of the fields and the energy densities (see \eqref{fri_km1}-\eqref{fri_km2}).    

\subsubsection{Pressureless matter coupled to both sectors.}  

We consider the equations of state
\begin{equation}
  p=0\quad\quad \tilde p=0,
\end{equation}
which corresponds to pressureless matter (or dust) coupled to both sectors.
\begin{figure}[h]
  \label{polvo}
  \centering
  \begin{subfigure}{.5\textwidth}
    \centering
    \includegraphics[width=.95\linewidth]{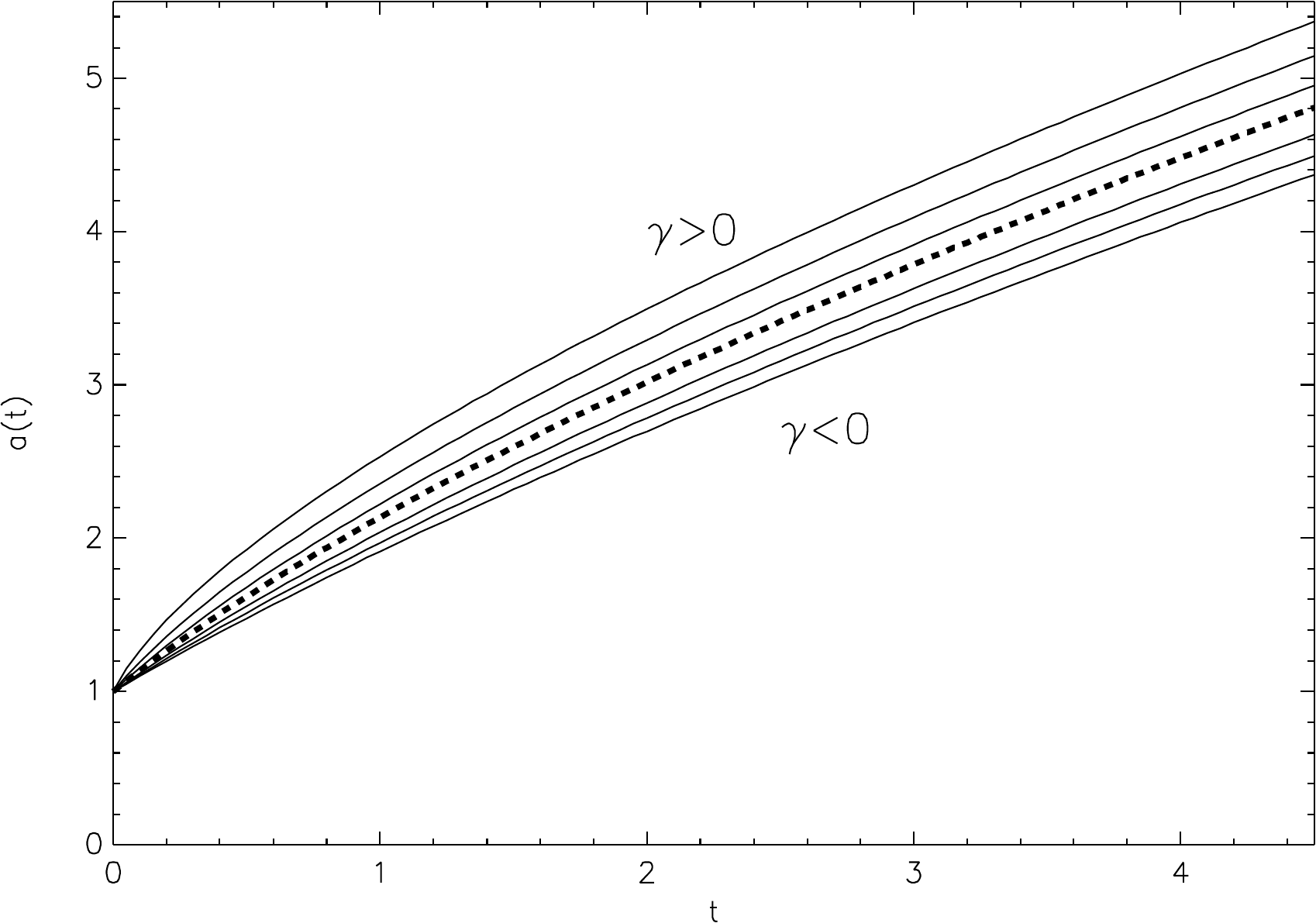}
    \caption{Evolution of the scale parameter $a(t)$.}
    \label{a_p}
  \end{subfigure}%
  \begin{subfigure}{.5\textwidth}
    \centering
    \includegraphics[width=.95\linewidth]{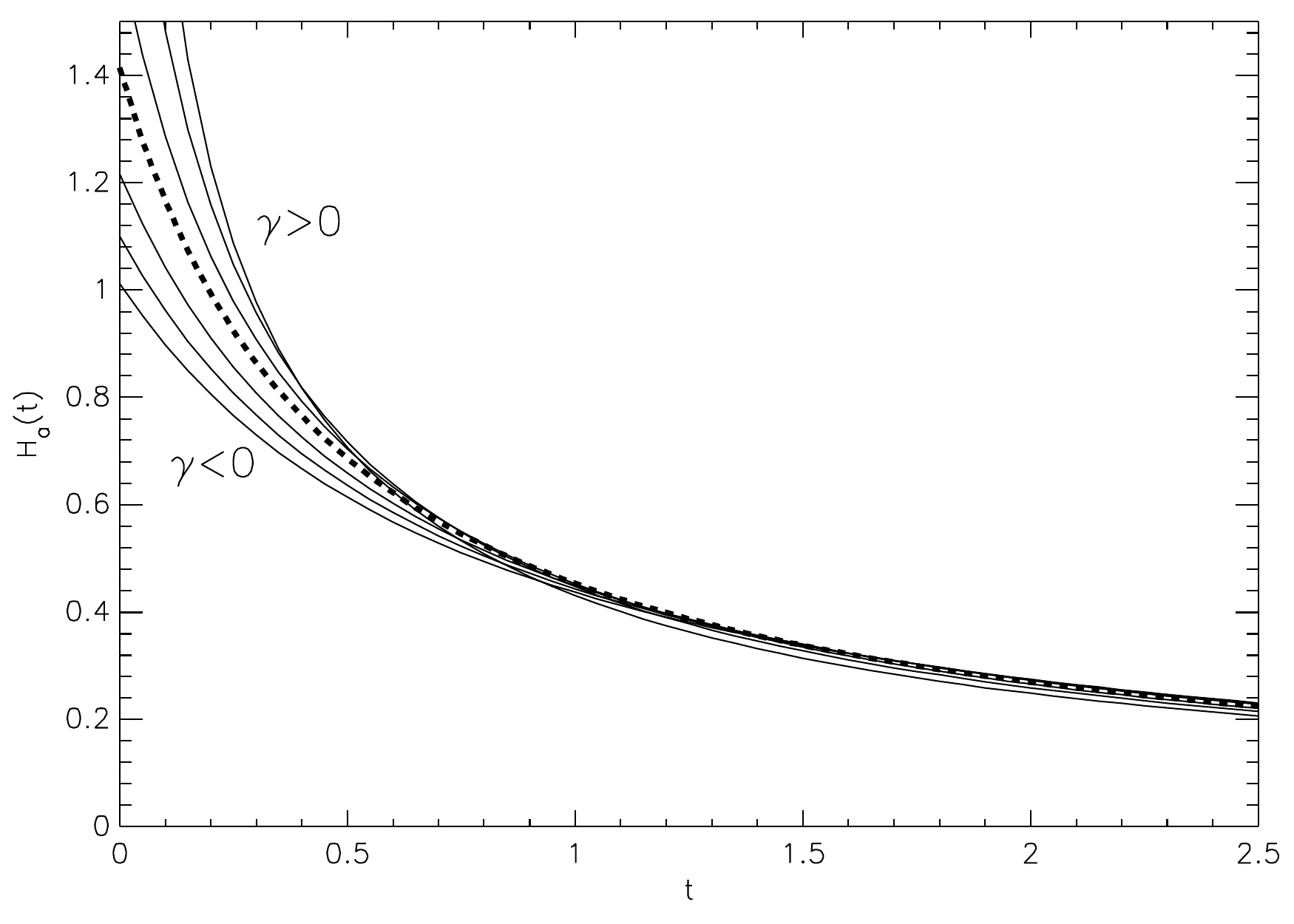}
    \caption{Evolution of the Hubble parameter $H_a(t)$.}
    \label{Ha_p}
  \end{subfigure}
  \begin{subfigure}{.5\textwidth}
    \centering
    \includegraphics[width=.95\linewidth]{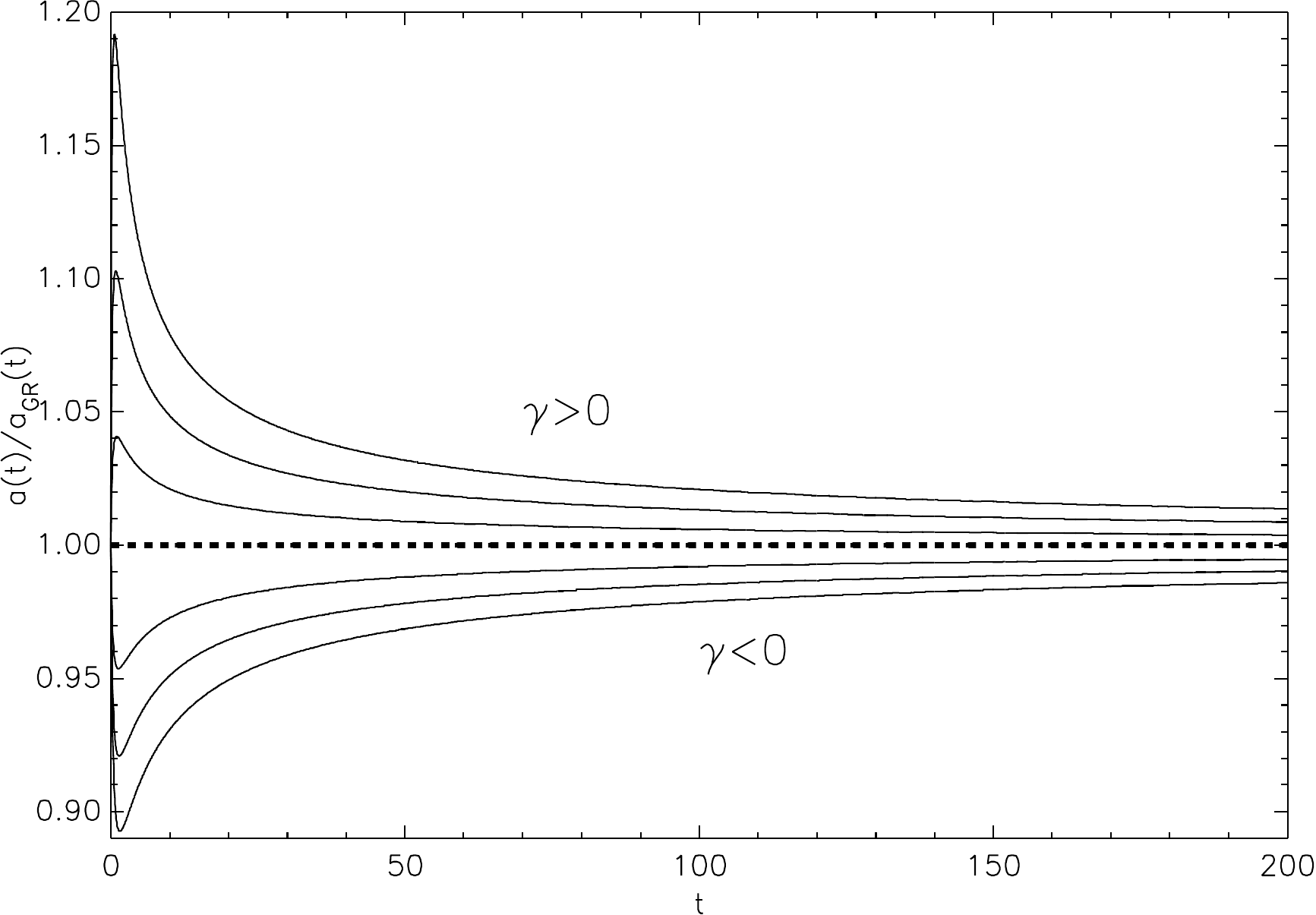}
    \caption{A comparison between the evolution of the scale parameter $a(t)$ with kinetic mixing and the standard GR evolution $a_{GR}(t)$ (coupled to dust).}
    \label{cuociente_p}
  \end{subfigure}
  \caption{Numerical solution of the field equations when pressureless matter is coupled to both sectors, for different values of the coupling constant ($\gamma=\{-0.95,-0.65,-0.35,0.25,0.55,0.85 \}$). Upper curves correspond to increasing values of $\gamma$. The dotted lines represent the standard GR solution. }
\end{figure}

The evolution of the scale parameter $a(t)$ for different values of $\gamma$ is displayed in figure \ref{a_p}. As it can be seen from the figure, the late time evolution of the scale factor $a(t)$ does not present new features compared to the standard cosmology solution $a_{GR}(t)$ (plotted as a dotted line) with the same initial conditions. Indeed, the Hubble parameter $H_a(t)=\frac{\dot a(t)}{a(t)}$ evolves similar to the standard case, as it can be noted from figure \ref{Ha_p}. Furthermore, for late times, the evolution of $a(t)$ is just a rescaling of $a_{GR}(t)$, as seen from figure \ref{cuociente_p}, which plots the evolution of $a(t)/a_{GR}(t)$. For each value of $\gamma$, this magnitude tends to a constant.   

\subsubsection{Radiation coupled to both sectors.}  

\begin{figure}[h]
  \label{radiacion}
  \centering
  \begin{subfigure}{.5\textwidth}
    \centering
    \includegraphics[width=.95\linewidth]{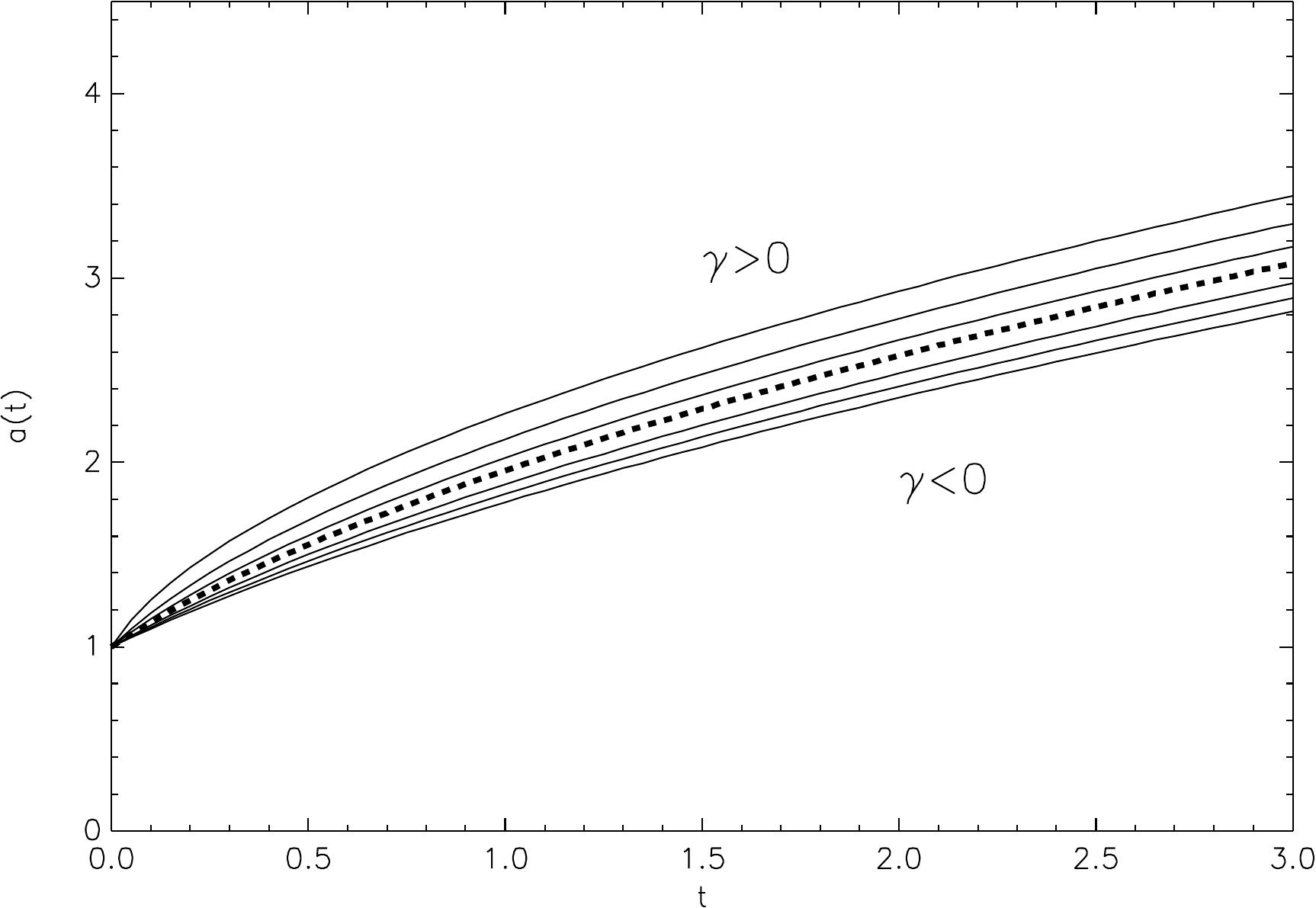}
    \caption{Evolution of the scale parameter $a(t)$.}
    \label{a_r}
  \end{subfigure}%
  \begin{subfigure}{.5\textwidth}
    \centering
    \includegraphics[width=.95\linewidth]{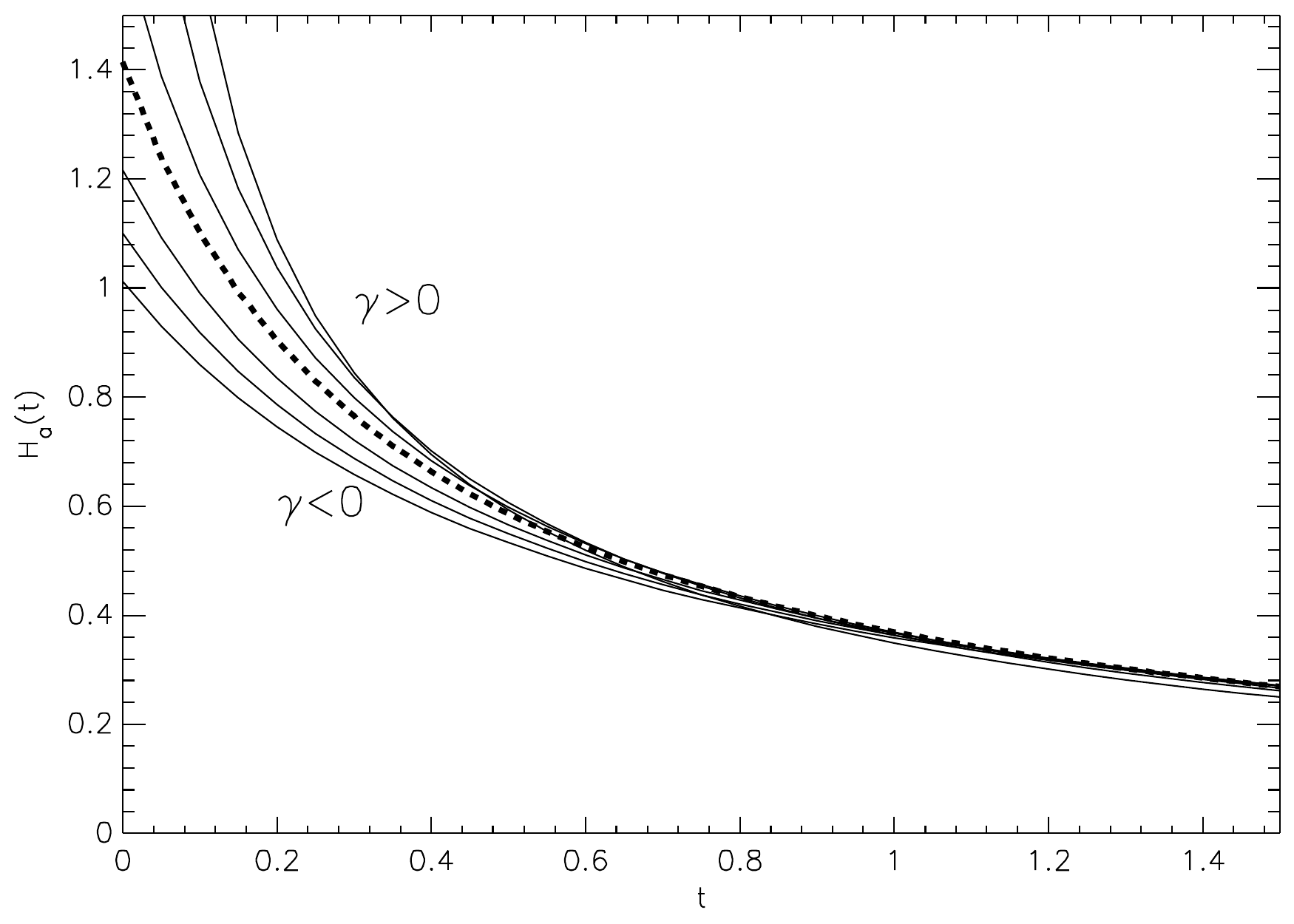}
    \caption{Evolution of the Hubble parameter $H_a(t)$.}
    \label{Ha_r}
  \end{subfigure}
  \begin{subfigure}{.5\textwidth}
    \centering
    \includegraphics[width=.95\linewidth]{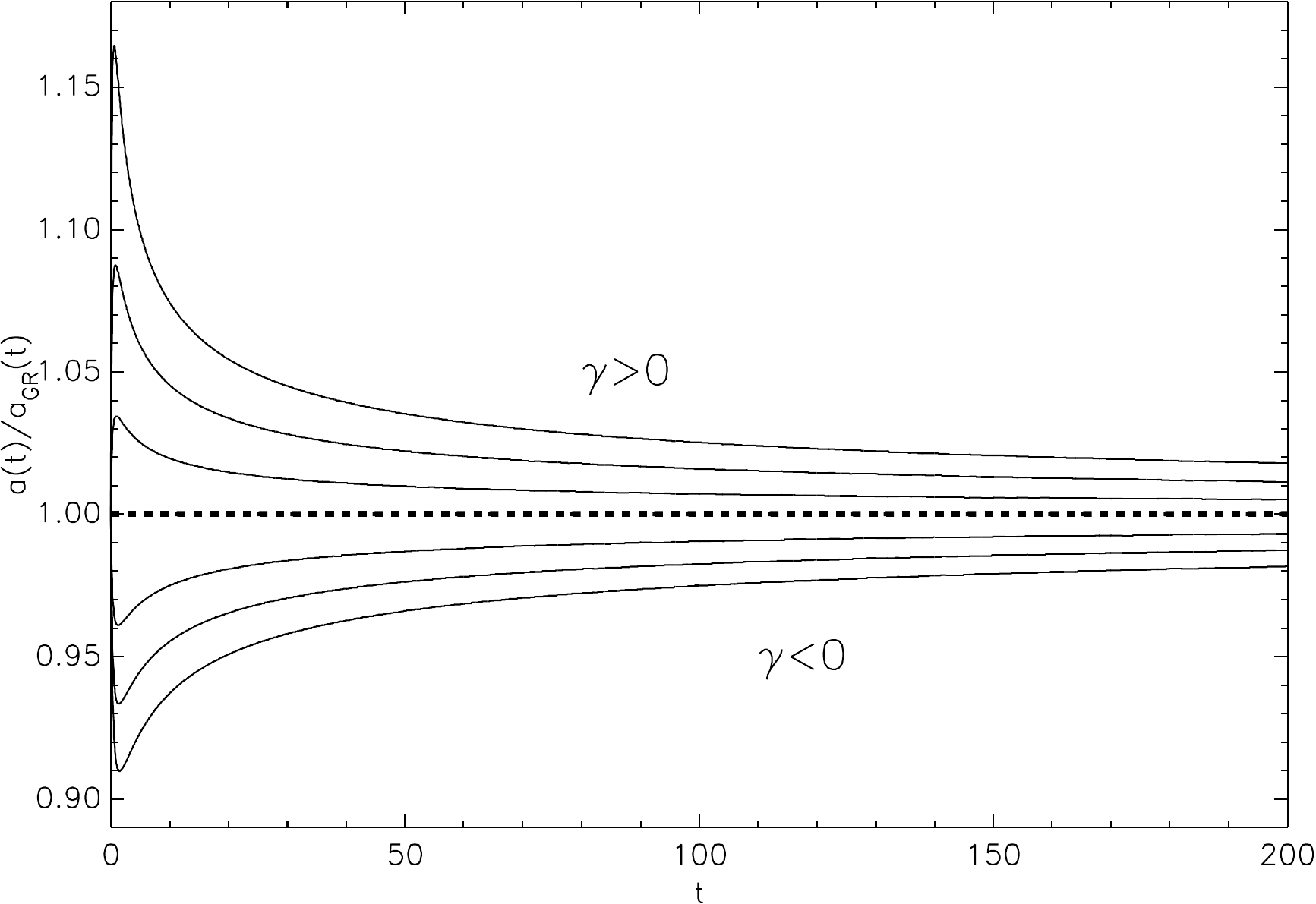}
    \caption{A comparison between the evolution of the scale parameter $a(t)$ with kinetic mixing and the standard GR evolution $a_{GR}(t)$ (coupled to radiation).}
    \label{cuociente_r}
  \end{subfigure}
  \caption{Numerical solution of the field equations when radiation is coupled to both sectors, for different values of the coupling constant ($\gamma=\{-0.95,-0.65,-0.35,0.25,0.55,0.85 \}$). Upper curves correspond to increasing values of $\gamma$. The dotted lines represent the standard GR solution. }
\end{figure}

Similar considerations can be made if we consider radiation coupled to both sectors of the theory. In this case, the appropriated equations of state are
\begin{equation}
  p=\frac{1}{3}\rho\quad\quad \tilde p=\frac{1}{3}\tilde \rho.
\end{equation}
Choosing the same initial conditions as in the previous case, the behavior of the scale factor is displayed in figure \ref{a_r} for different values of $\gamma$. The Hubble parameter is plotted in figure \ref{Ha_r}, while the ratio $a(t)/a_{GR}(t)$ is in figure \ref{cuociente_r}. Just like the previous case, the late time evolution is similar to standard cosmology with radiation.

\subsubsection{Pressureless matter coupled to $a$-sector, radiation coupled to $b$-sector}

\begin{figure}[p]
\label{polvo_radiacion}
\centering
\begin{subfigure}{.5\textwidth}
  \centering
  \includegraphics[width=.95\linewidth]{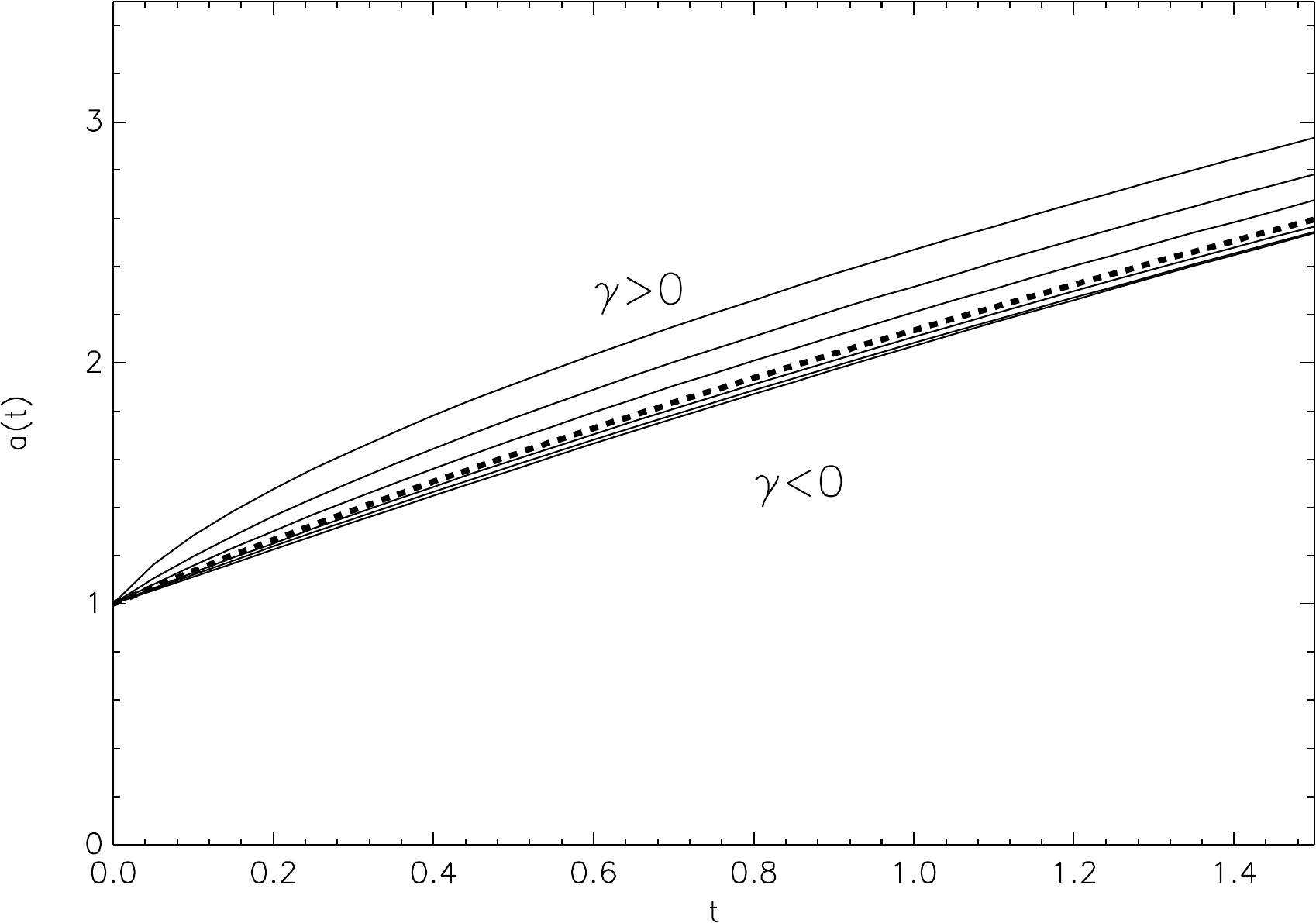}
  \caption{Evolution of the scale parameter $a(t)$.}
  \label{a_pyr}
\end{subfigure}%
\begin{subfigure}{.5\textwidth}
  \centering
  \includegraphics[width=.95\linewidth]{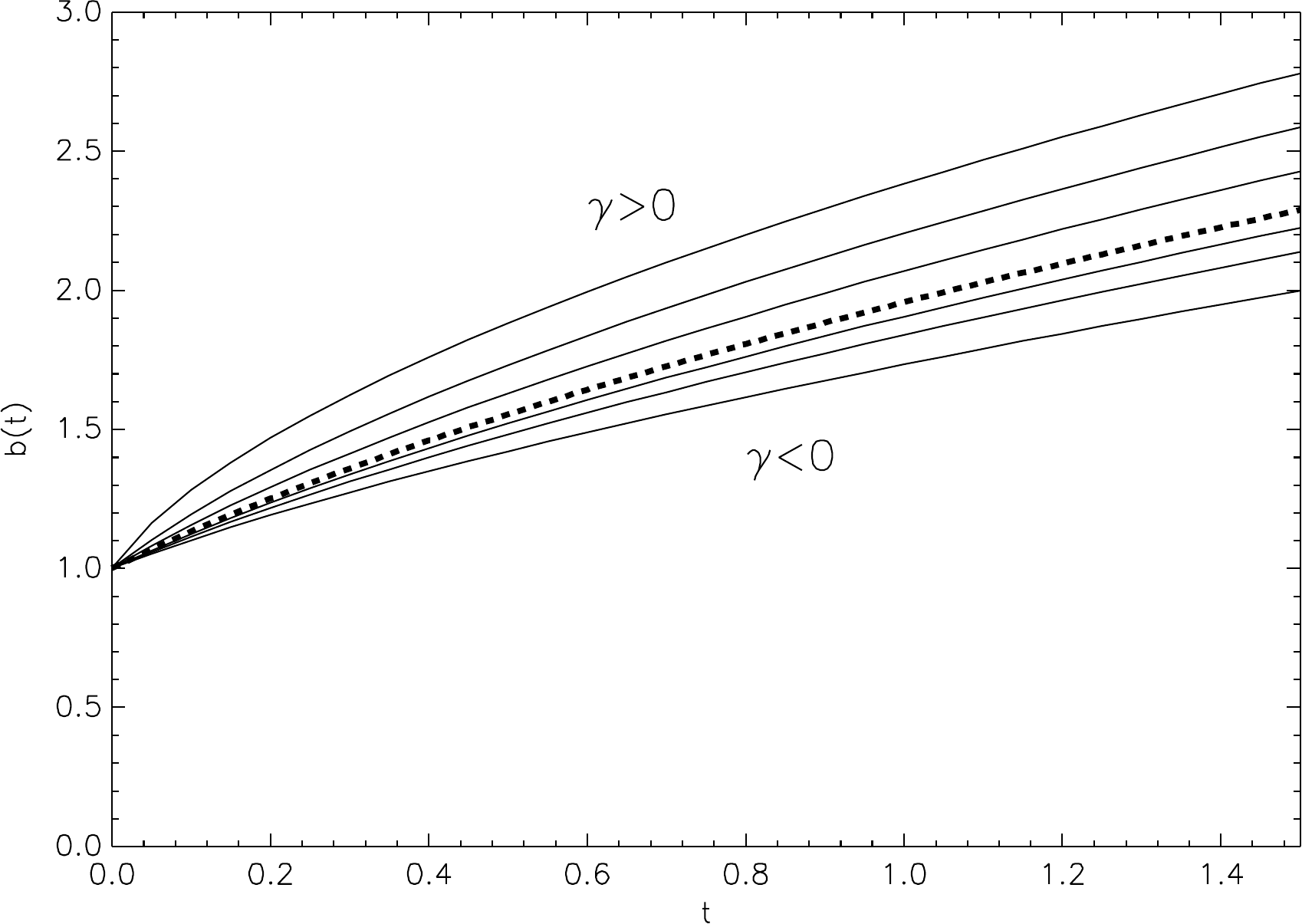}
  \caption{Evolution of the scale parameter $b(t)$.}
  \label{b_pyr}
\end{subfigure}

\begin{subfigure}{.5\textwidth}
  \centering
  \includegraphics[width=.95\linewidth]{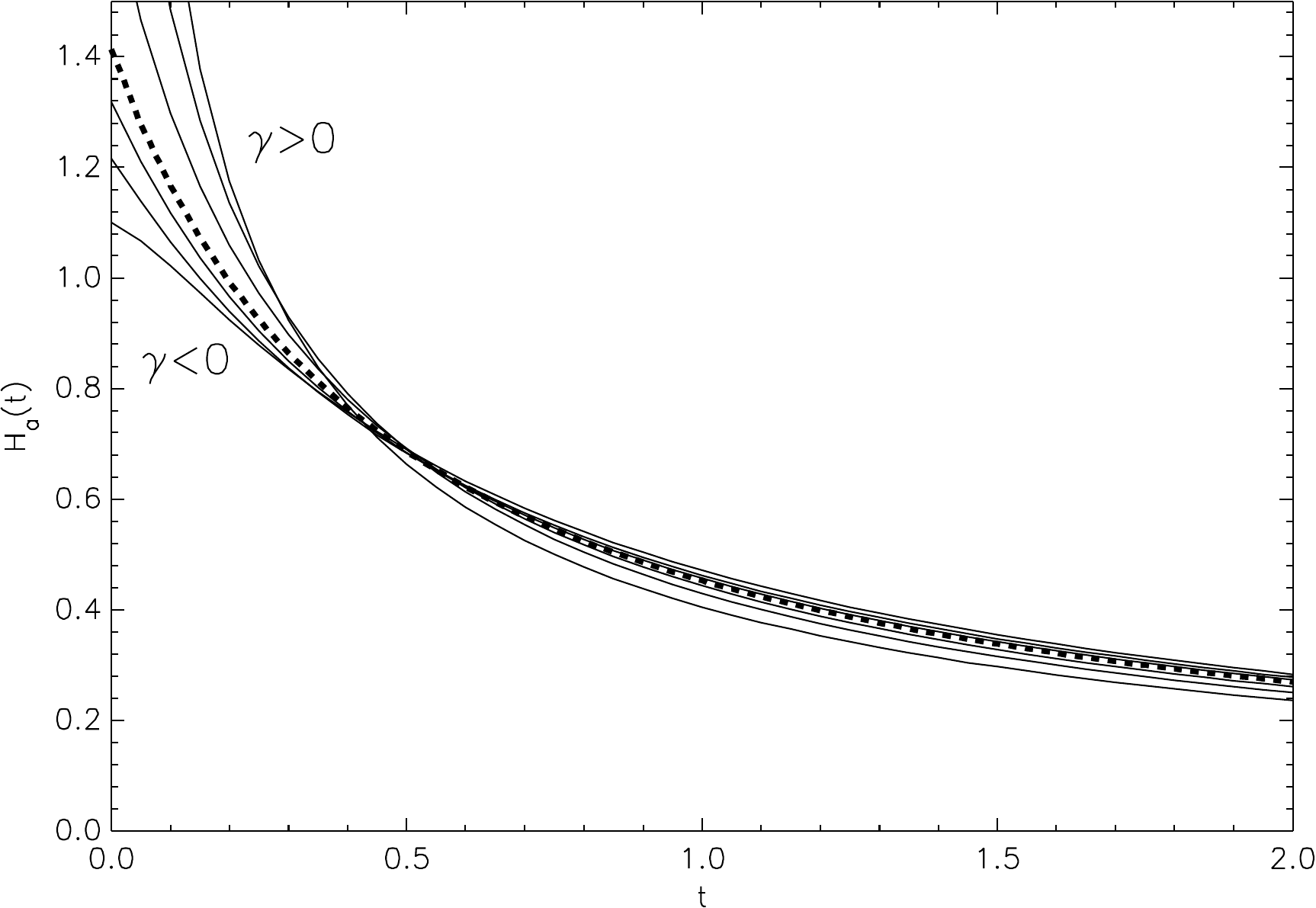}
  \caption{Evolution of the Hubble parameter $H_a(t)$.}
  \label{Ha_pyr}
\end{subfigure}%
\begin{subfigure}{.5\textwidth}
  \centering
  \includegraphics[width=.95\linewidth]{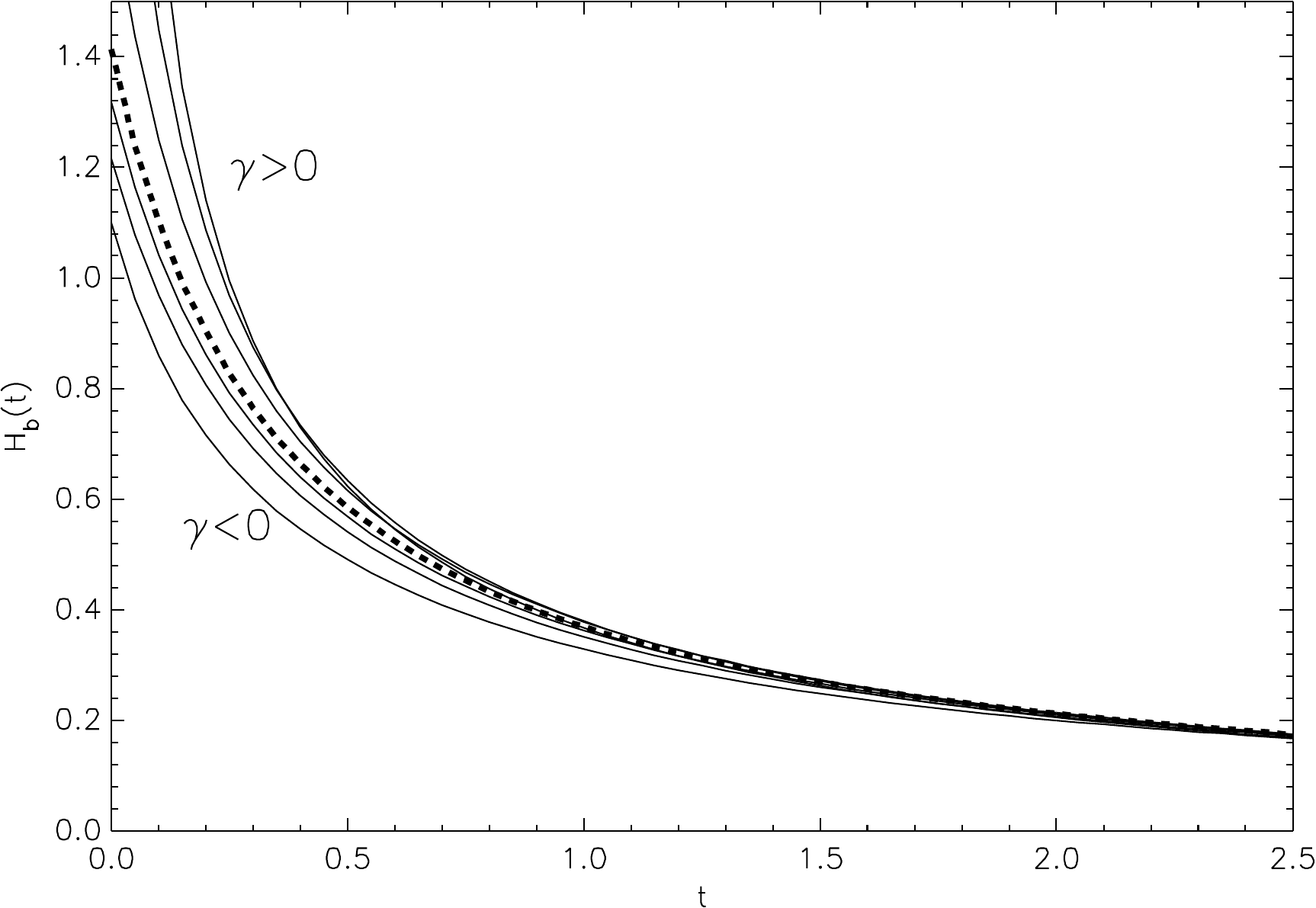}
  \caption{Evolution of the Hubble parameter $H_b(t)$.}
  \label{Hb_pyr}
\end{subfigure}

\begin{subfigure}{.5\textwidth}
  \centering
  \includegraphics[width=.95\linewidth]{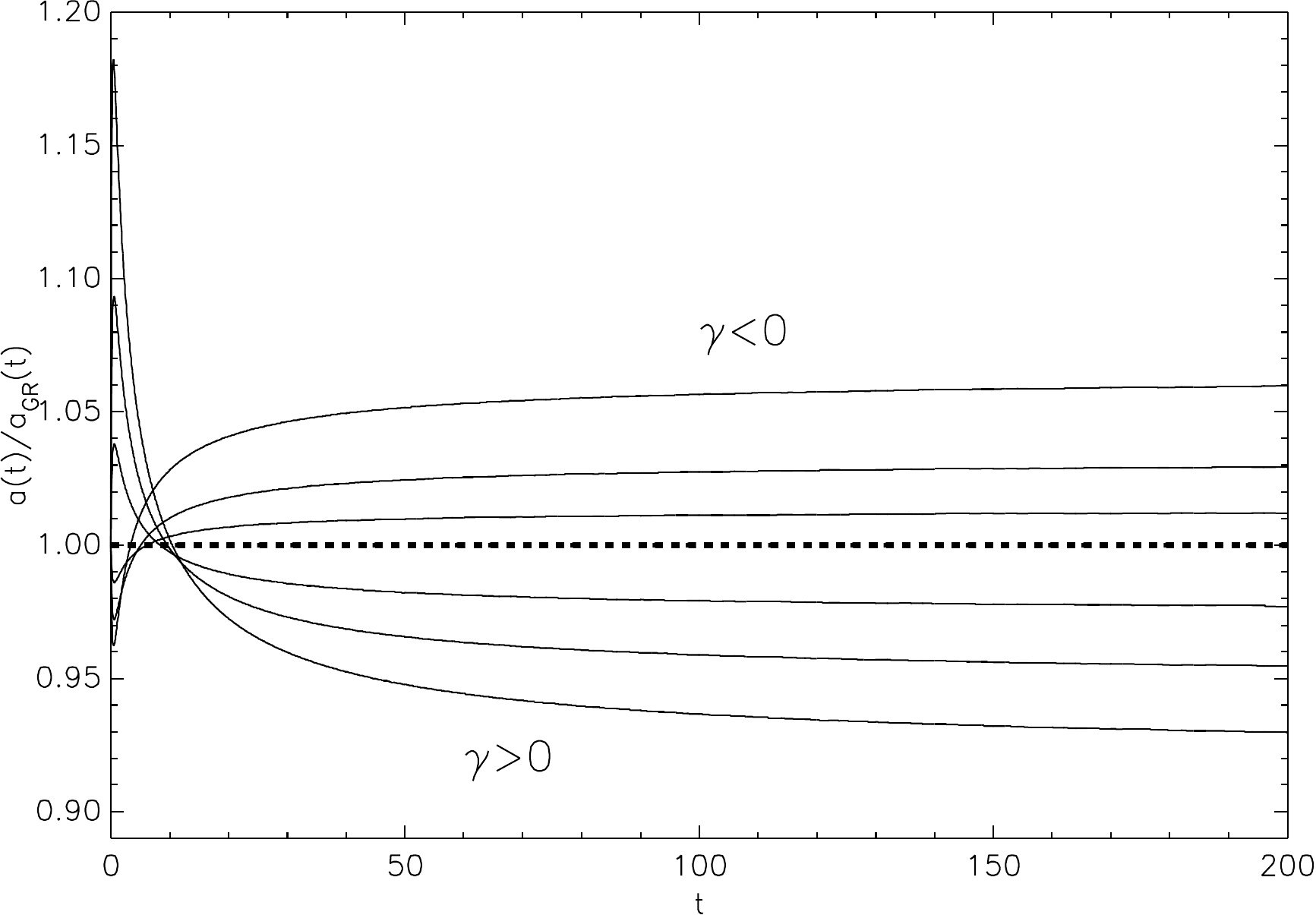}
  \caption{A comparison between the evolution of the scale parameter $a(t)$ with kinetic mixing and the standard GR evolution $a_{GR}(t)$ (coupled to dust).}
  \label{cuocientea_pyr}
\end{subfigure}%
\begin{subfigure}{.5\textwidth}
  \centering
  \includegraphics[width=.95\linewidth]{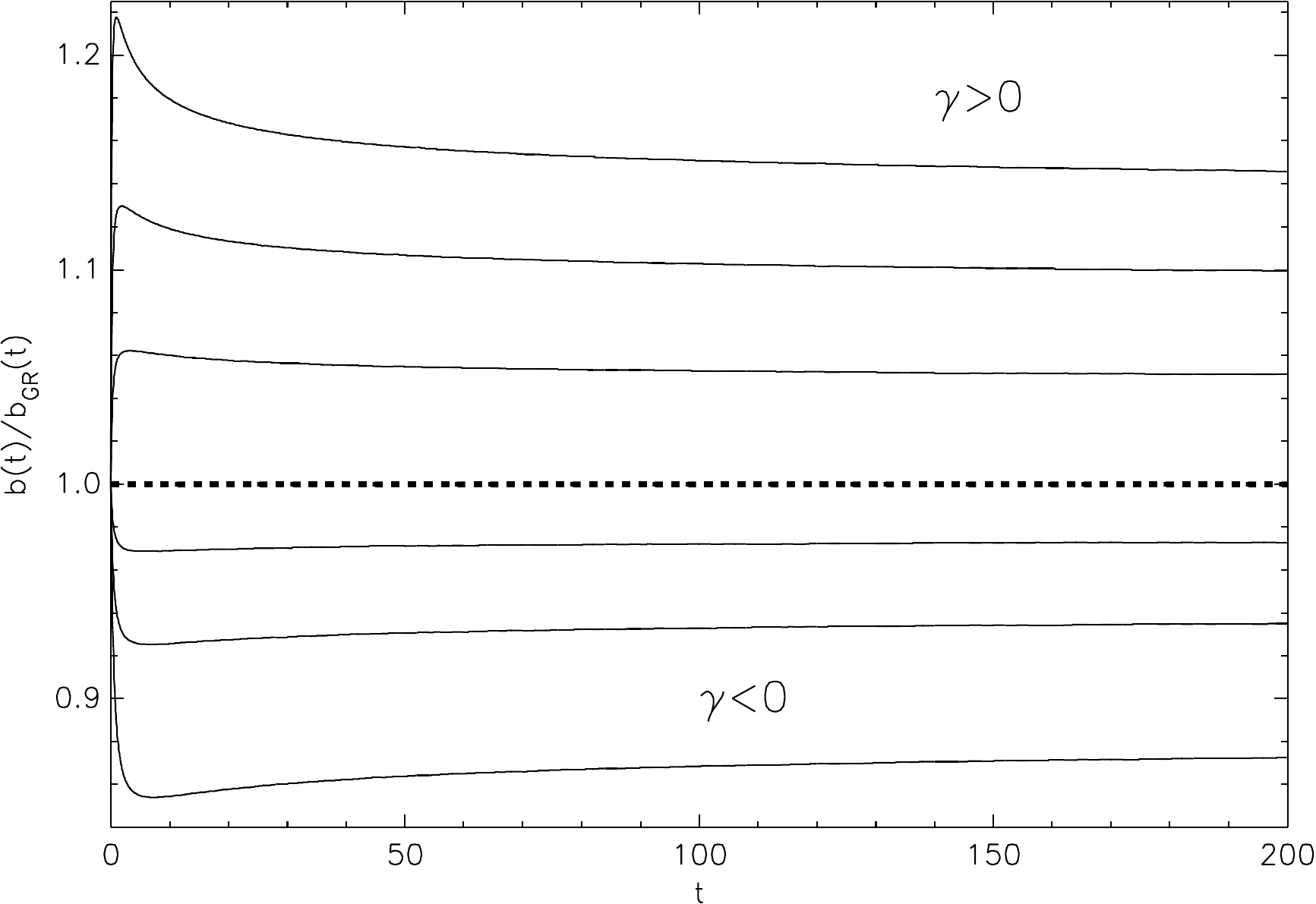}
  \caption{A comparison between the evolution of the scale parameter $b(t)$ with kinetic mixing and the standard GR evolution $b_{GR}(t)$ (coupled to radiation).}
  \label{cuocienteb_pyr}
\end{subfigure}

\caption{Numerical solution of the field equation when dust is coupled the $a$-sector, while radiation couples to the $b$-sector, for different values of the coupling constant ($\gamma=\{-0.65,-0.35,-0.15,0.3,0.6,0.9 \}$). Upper curves correspond to increasing values of $\gamma$. The dotted lines represent the standard GR solutions. }
\end{figure}

In the two above cases, both sectors are treated in a symmetric way. Now we couple different kind of matter (with different equations of state) to each sector and show that the above conclusion remain, i.e., the late time evolution of the scale factors is similar to standard cosmology. For definiteness, we couple pressureless matter to the $a$-sector and radiation to the $b$-sector. In this case, the pertinent equations of state are
\begin{equation}
  p=0\quad\quad \tilde p=\frac{1}{3}\tilde \rho.
\end{equation}
The same plots as in the above sections are displayed. The evolution of $a(t)$ and $b(t)$ (figures \ref{a_pyr} and \ref{b_pyr}), the Hubble parameters $H_a(t)$ and $H_b(t)$ (figures \ref{Ha_pyr} and \ref{Hb_pyr}) and the ratios $a(t)/a_{GR}(t)$ and $b(t)/b_{GR}(t)$ (figures \ref{cuocientea_pyr} and \ref{cuocienteb_pyr}) show no particular new features than those from general relativity.

\subsubsection{Cosmological constant coupled to $a$-sector, pressureless matter coupled to $b$-sector}

\begin{figure}[p]
\label{lambda_polvo}
\centering
\begin{subfigure}{.5\textwidth}
  \centering
  \includegraphics[width=.95\linewidth]{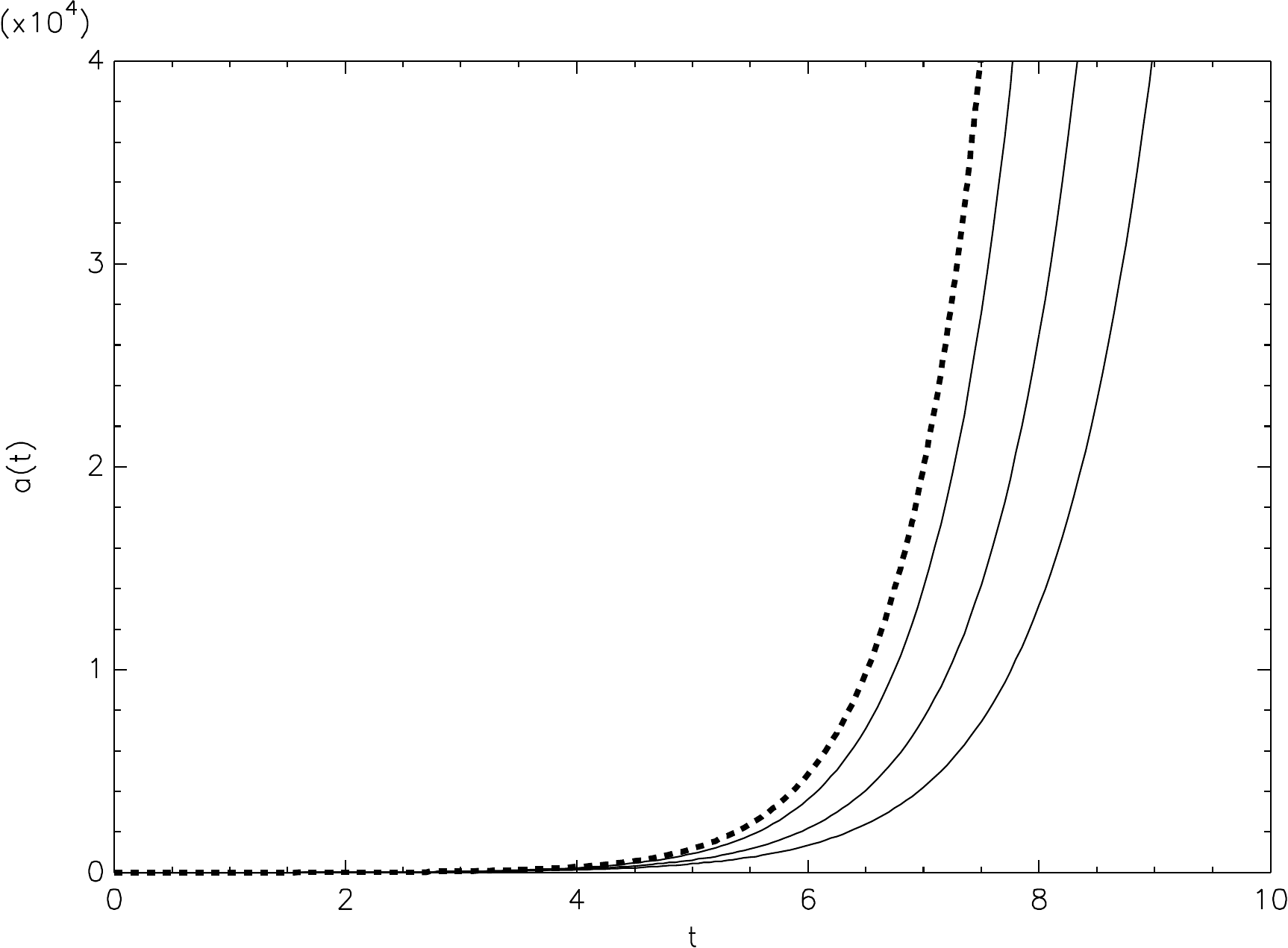}
  \caption{Evolution of the scale parameter $a(t)$.}
  \label{a_lyp}
\end{subfigure}%
\begin{subfigure}{.5\textwidth}
  \centering
  \includegraphics[width=.95\linewidth]{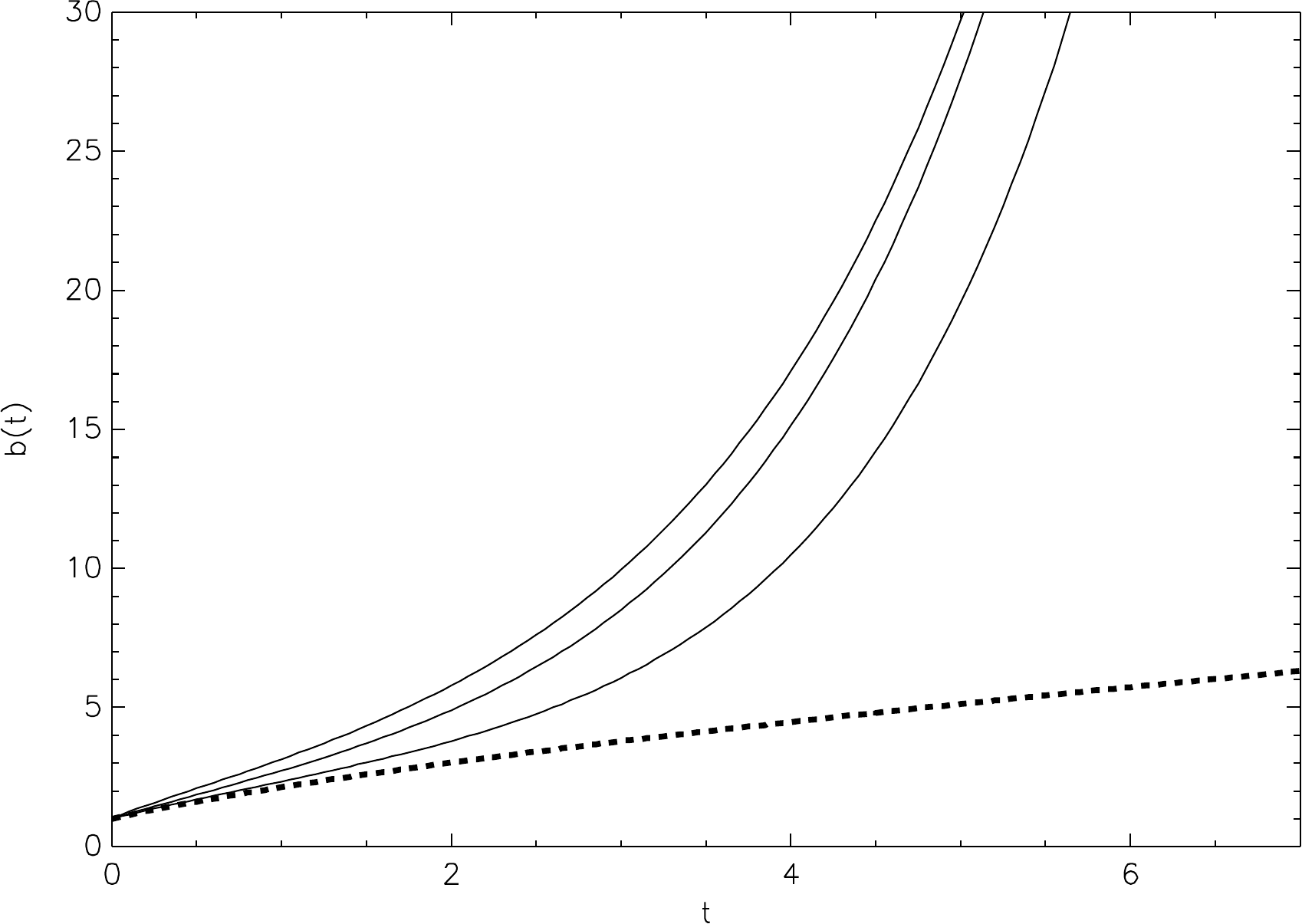}
  \caption{Evolution of the scale parameter $b(t)$.}
  \label{b_lyp}
\end{subfigure}

\begin{subfigure}{.5\textwidth}
  \centering
  \includegraphics[width=.95\linewidth]{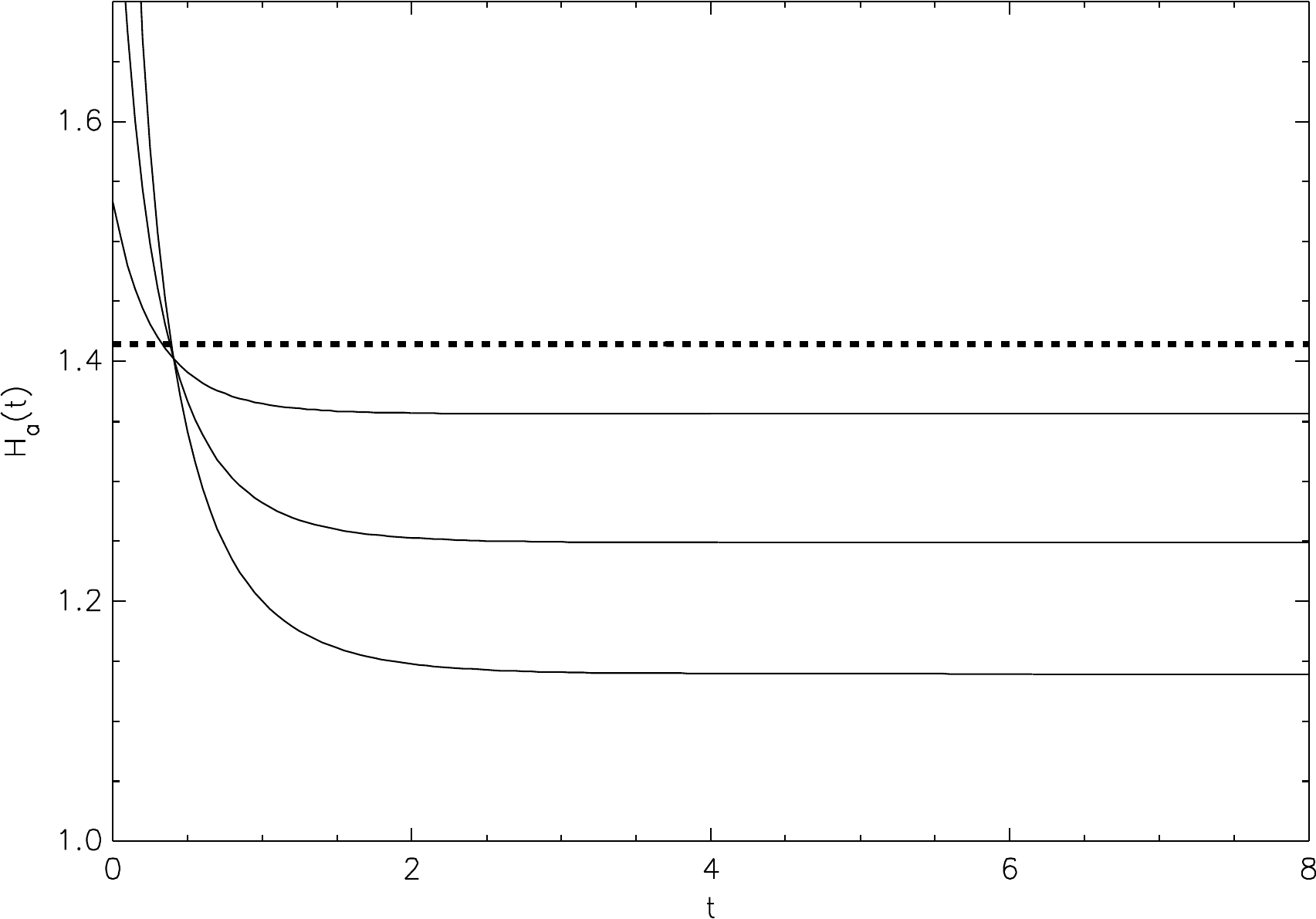}
  \caption{Evolution of the Hubble parameter $H_a(t)$.}
  \label{Ha_lyp}
\end{subfigure}%
\begin{subfigure}{.5\textwidth}
  \centering
  \includegraphics[width=.95\linewidth]{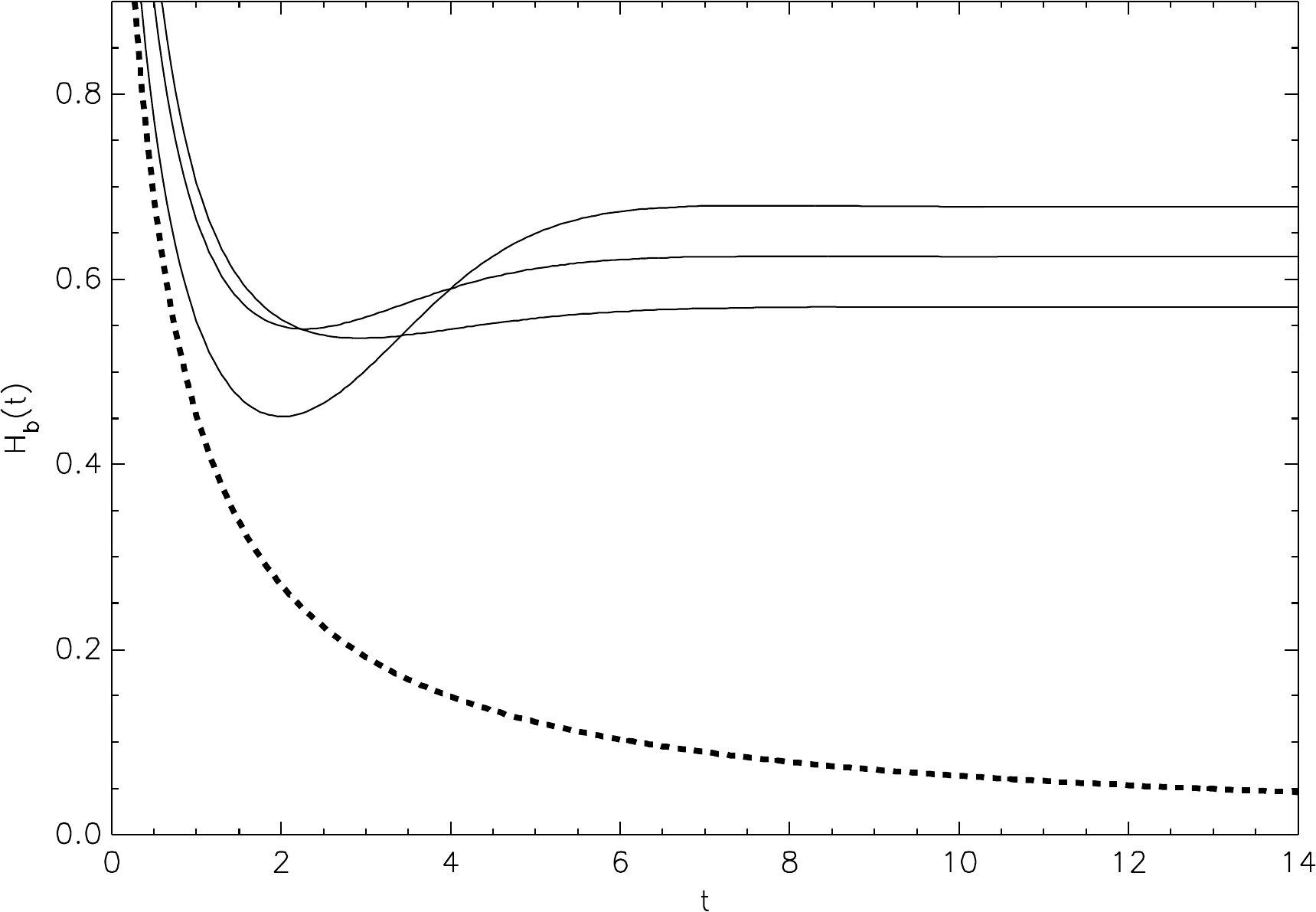}
  \caption{Evolution of the Hubble parameter $H_b(t)$.}
  \label{Hb_lyp}
\end{subfigure}

\begin{subfigure}{.5\textwidth}
  \centering
  \includegraphics[width=.95\linewidth]{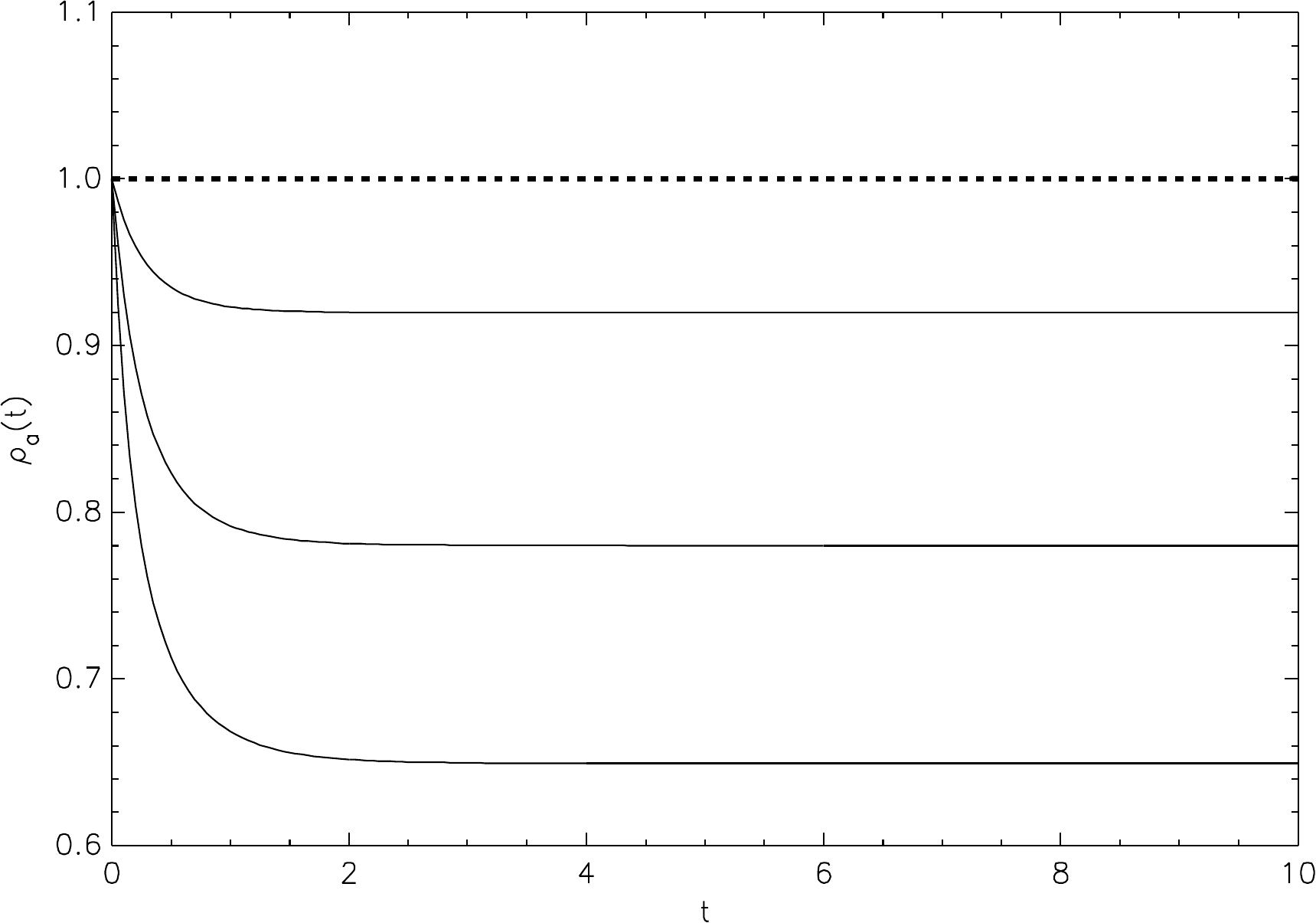}
  \caption{Evolution of the energy density $\rho(t)$ in the $a$-sector.}
  \label{rhoa_lyp}
\end{subfigure}%
\begin{subfigure}{.5\textwidth}
  \centering
  \includegraphics[width=.95\linewidth]{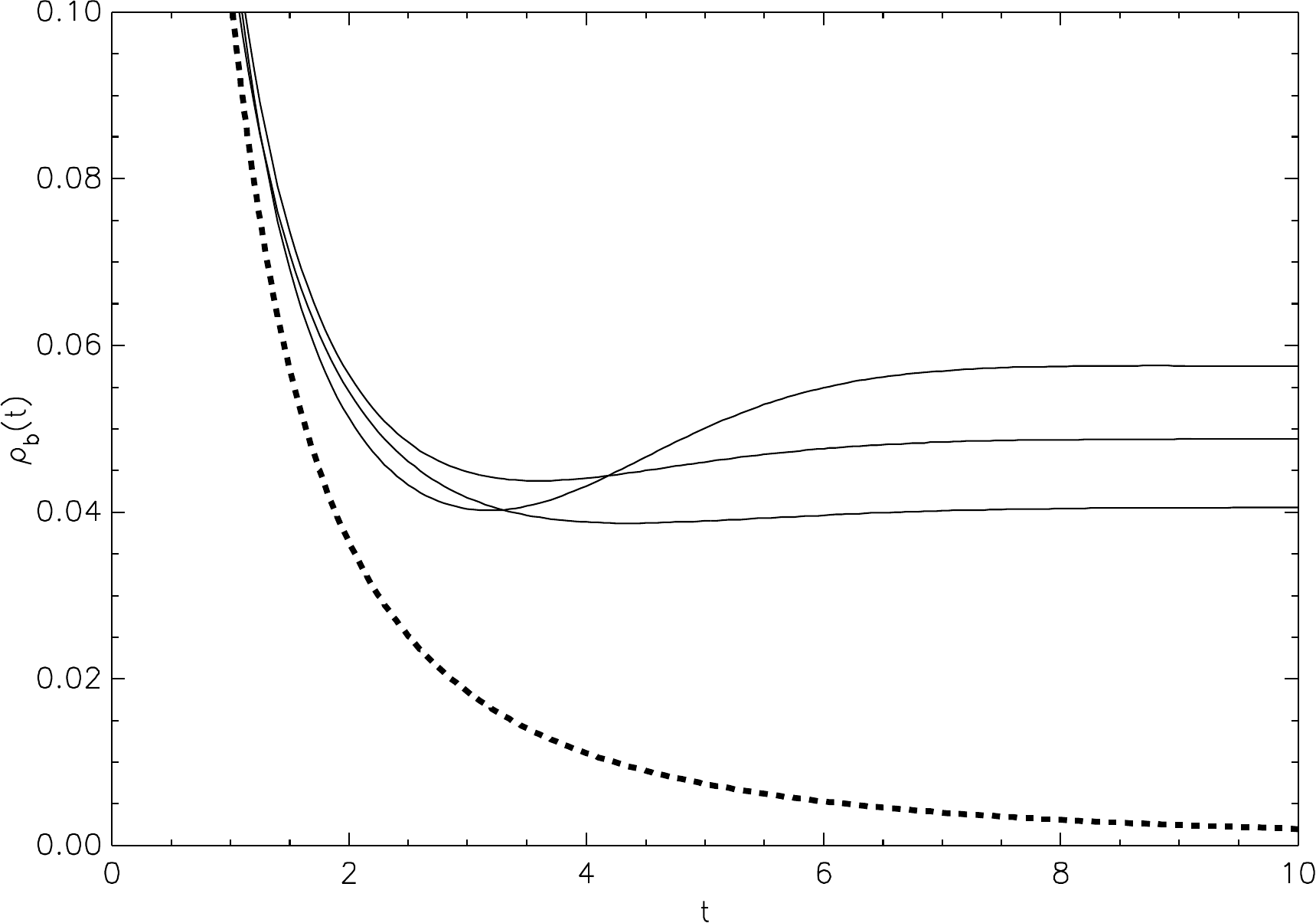}
  \caption{Evolution of the energy density $\tilde \rho(t)$ in the $b$-sector.}
  \label{rhob_lyp}
\end{subfigure}

\caption{Numerical solution of the field equation when a cosmological constant couples to the $a$-sector, while dust couples to the $b$-sector, for different values of the coupling constant ($\gamma=\{0.15,0.3,0.75\}$). The closest curves to the dotted line correspond to smaller values of $\gamma$. The dotted lines represent the standard GR solution. }
\end{figure}

As opposite to the above cases, when coupling a cosmological constant, the Kinetic Mixing plays an important role. To see this, we choose equations of state describing a constant energy density in the $a$-sector and dust in the $b$-sector. These are given by
 \begin{equation}
  p=-\rho\quad\quad \tilde p=0.
\end{equation}
We can see from figures \ref{a_lyp} and \ref{b_lyp} that the behavior of the scale factors $a(t)$ and $b(t)$ differ drastically from the GR case. In figure \ref{a_lyp} we can see the evolution of $a(t)$ for different values of $\gamma$. In all cases the growth is exponential, just as in GR. Nonetheless, for increasing values of $\gamma$, the velocity of the expansion is slower, which indicates that kinetic mixing produces a rescaling of the cosmological constant. Another way to see this effect is with figure \ref{rhoa_lyp}, where the evolution of the $a$-sector energy density $\rho$ is displayed. In ordinary GR with cosmological constant, the energy density remains constant in time, while for $\gamma \neq 0$ the energy density drops to a lower constant value. Similar observation can be made in figure \ref{Ha_lyp} from where we can see that the Hubble parameter is constant, but smaller than standard GR, when $\gamma$ is turned on. 

On the other hand, the evolution of the $b(t)$ parameter becomes exponential, although there is only dust coupled in this side (see figure \ref{b_lyp}). Note from figure \ref{rhob_lyp}  that the energy density $\tilde \rho$ becomes constant for late times, as opposite to GR, where it drops to zero. Similarly, the Hubble parameter becomes constant in time (figure \ref{Hb_lyp}). This indicates that the $b$-sector of the theory feels an effective cosmological constant due to the presence of the kinetic mixing.

Summarizing, in this setup, the kinetic mixing has two effects: it shields the cosmological constant in the $a$-sector and turns on an effective cosmological constant in the $b$-sector.

\section{Discussion and conclusions}\label{conclusions} 

Hoping to understand how to construct a kinetic mixing mechanism for gravity, we have turned our attention to simple models that are not fully covariant theories but rather have a time reparametrization invariance, which can be regarded as a ``simpler'' case of covariance. The first system considered is composed of two relativistic particles interacting with each other, while the second system is a cosmological model. Both systems are related to gravity; the Lagrangian for the massive relativistic particle \eqref{relN} can be regarded as a one-dimensional theory of gravity interacting with scalar fields, while the cosmological model considered in \eqref{lan} can be easily obtained by a minisuperspace construction starting from General Relativity. In both cases, the kinetic mixing procedure has given similar results.     

For the interacting relativistic particles \eqref{part1}, the system can be decoupled by means of an appropriated non-local redefinition of  fields, leaving two non-interacting particles. However, as a result of this redefinition, the mass of one the particles gets rescaled by a factor $\sqrt{1-\gamma^2}$, where $\gamma$ is the coupling constant of the kinetic mixing. It is important to notice that one of the particle must be chosen massless since including mass terms for both particles would spoil the diagonalization procedure. This rescaling of the mass is similar to what happens with kinetic mixing in electrodynamics \eqref{lag}, where it is the electric charge that gets rescaled, allowing for phenomenological searches of this hidden interaction. The diagonalization procedure also makes explicit the symmetries of the system \eqref{part1}. Since it describes two independent particles, the system possesses two independent time reparametrization symmetries \eqref{symmetries}. This also resembles the electromagnetic case where, after the diagonalization, the two $U(1)$ gauge symmetries are explicit. 

For the interacting cosmological models \eqref{min1} it is more difficult to obtain conclusions since the theory is non-linear. However, it is straightforward to prove that, similarly to the particles case, this system can be decoupled by means of a non-local redefinition of the fields, leaving explicitly two gauge symmetries (time reparametrizations). 

In order to study the consequences of the kinetic mixing in the cosmological setup, it is easier to study the evolution of the fields through numerical integration of the equations of motion. Several scenarios can be studied, depending on which kind of matter is coupled to the different sectors of the theory. In figures 1, 2  and 3 it is displayed the evolution of the fields with time, when dust, radiation or a combination of both are present in the system. It can be seen from the figures that no significant difference occurs when the kinetic mixing is turned on (compared to the standard cosmology evolution with the same initial conditions). For the three cases, the late time evolution of the scale factors tends to a rescaling of the standard GR solution. 

It is worthwhile to mention that the evolution of the scale factors is always decelerating in the cases of figure 1 and 2, fact that can be easily checked numerically. However, in the case of figure 3 (dust coupled to the $a$-sector, radiation in the other) with negative $\gamma$, the initial evolution of the scale factor $a(t)$ presents a short period of acceleration, whose extension in time increases as the value of $\gamma$ shifts away from zero. Although this is not a period where the Hubble parameter $H_a$ is constant, it may be useful to accommodate a period of inflation. Further studies in this direction will be left for future work.        

The situation is very different when a cosmological constant is coupled to the system. In figure 4 it is plotted the evolution of the system when a cosmological constant is coupled to the $a$ side and dust on the $b$ side. In this setup, the kinetic mixing has two effects. First, the value of the energy density in the $a$-sector (constant in the case of standard GR) drops a to smaller constant value. This indicates that the cosmological constant in the $a$ side gets screened to a smaller value, similar to the effect of the rescaling of the mass in the relativistic particles example. On the other side, the $b$-sector of the theory evolves exponentially, attaining an effective cosmological constant, indicating that the kinetic mixing allows the $b$ side to feel the cosmological constant coupled in the $a$ side. As mentioned above, a similar observation can be made for the relativistic particles, when one particle interacts with the matter coupled to the other one. Note that the value of the constant energy density in the $b$-sector is smaller than the $a$-sector.

The rescaling observed numerically may help in understanding the smallness of the measured cosmological constant. Recall that the observed value for it is about 120 order of magnitude smaller than the prediction from QFT. A process such as kinetic mixing might help to alleviate the problem. However, note from figures \ref{rhoa_lyp} and \ref{rhob_lyp} that values of $\gamma$ of the order $1-10^{-1}$ generates values for the energy density of $10^{-1}$ in the $a$-sector and $10^{-2}$ in the $b$-sector (in units used in the paper). It is unclear to us whether the observed rescaling might correct 120 orders of magnitude without introducing an unnaturally large value for $\gamma$. 

Another way to interpret the result is by considering that our universe is coupled to the $b$-sector of the theory, containing only normal matter, and the cosmological constant measured by observations might be due to a kinetic mixing with a hidden sector.                  

Although the examples mentioned here are just simple cases of fully covariant theories, the kinetic mixing mechanism developed for the cosmological case still leaves much work to be done. Further study must be done on how this modifies the inflation scenario and to understand the evolution of cosmological perturbation in this context.

\acknowledgements 
This work was supported by FONDECYT/Chile grants 1130020 (J.G.), 7912010045 and 11130083 (M.P.)

\section*{Appendix}

Here we present the infinitesimal symmetry transformations of the action \eqref{action11}, corresponding to the two independent time reparametrizations. 

Recall that using the field redefinition \eqref{redef11}
\begin{equation}\label{redef12}
   n\equiv\frac{N}{1-\frac{\gamma^2}{ab}} , \quad\quad \dot c\equiv \dot b -\gamma\sqrt{\frac{M}{N}}\frac{\dot a}{b}, \quad\quad m\equiv M \frac{c}{b} ,
\end{equation}
 the action \eqref{action11} becomes diagonalized
\begin{equation}
   L= -3\frac{a \,{\dot a}^2}{n} - 3\frac{c \,{\dot c}^2}{ m}.
\end{equation}
In this setting, both symmetries can be easily written as
\begin{align}\label{trans11}
    \delta a &= \epsilon_1 \dot{a}, \quad\quad \delta n = \frac{d}{dt}(\epsilon_1 n),\nonumber\\
    \delta c &= \epsilon_2 \dot{c}, \quad\quad \delta m = \frac{d}{dt}(\epsilon_2 m),
\end{align}
where $\epsilon_1$ and $\epsilon_2$ are two arbitrary functions of time.

To find the transformations in terms of the original fields, we perform a variation of the relations \eqref{redef12}, yielding
\begin{align}
  \delta n &= \frac{1}{1-\frac{\gamma^2}{ab}}\delta N -\frac{\gamma^2 N }{a^2b\left(1-\frac{\gamma^2}{ab}\right)^2}\delta a - \frac{\gamma^2 N }{ab^2\left(1-\frac{\gamma^2}{ab}\right)^2}\delta b,\\
\delta m&=\frac{c}{b}\delta M + \frac{M}{b}\delta c -\frac{Mc}{b^2}\delta b,\\
\delta \dot c &=\delta \dot b -\gamma \sqrt{\frac{M}{N}}\frac{1}{b}\delta \dot a+\gamma \sqrt{\frac{M}{N}}\frac{\dot a}{b^2}\delta b-\gamma \frac{\dot a}{2b\sqrt{MN}}\delta M+\gamma \sqrt{\frac{M}{N^3}}\frac{\dot a}{2b}\delta N.
\end{align}
Using the transformations \eqref{trans11}, and with the aid of the relations \eqref{redef12}, the above equations can be solved to find $\delta b$, $\delta N$, and $\delta M$ in terms of the fields and the parameters $\epsilon_1$ and $\epsilon_2$ (The transformation $\delta a$ is already given in \eqref{trans11}). Of course, such a procedure cannot be done algebraically, since the relation between $c$ and the other fields is non-local in time, which implies non-local field transformations.

\end{document}